\documentclass[12pt]{article}

\usepackage{latexsym, graphicx, amssymb, amsmath, mathrsfs, physics, bm}
\usepackage{here}
\usepackage{cite,color}
\usepackage{hyperref}
\usepackage{comment}
\usepackage{float}

\newcommand{\im}{\mathrm{i}\mkern1mu}

\numberwithin{equation}{section}

\begin{document}
\begin{titlepage}

\begin{flushright} 
RIKEN-iTHEMS-Report-25, UTHEP-801, KEK-TH-2355
\end{flushright}

\vspace{1cm}

\begin{center}
{\bf \large
  On the validity of the complex Langevin method
  near the deconfining phase transition in QCD at finite density 
  }
\end{center}
\vspace{1cm}

\renewcommand{\thefootnote}{\fnsymbol{footnote}}

\begin{center}
Shoichiro T{\sc sutsui}$^{a}$\footnote{E-mail address : shoichiro.tsutsui@a.riken.jp},
Yuhma A{\sc sano}$^{b}$\footnote{E-mail address : asano@het.ph.tsukuba.ac.jp},
Yuta I{\sc to}$^{c}$\footnote{E-mail address : y-itou@tokuyama.ac.jp},
Hideo M{\sc atsufuru}$^{de}$\footnote{E-mail address : hideo.matsufuru@kek.jp},
Yusuke N{\sc amekawa}$^{f}$\footnote{E-mail address : namekawa@fukuyama-u.ac.jp},
Jun N{\sc ishimura}$^{eg}$\footnote{E-mail address : jnishi@post.kek.jp},
Shinji S{\sc himasaki}$^{h}$\footnote{E-mail address : shimasaki.s@gmail.com},
Asato T{\sc suchiya}$^{i}$\footnote{E-mail address : tsuchiya.asato@shizuoka.ac.jp}

\vspace{1cm}

$^a${\it Interdisciplinary Theoretical and Mathematical Sciences Program (iTHEMS),
         RIKEN,
         Saitama 351-0198, Japan} 

$^b${\it Institute of Pure and Applied Sciences,
         University of Tsukuba,
	 Ibaraki 305-8577, Japan} 

$^c${\it National Institute of Technology,
         Tokuyama College,
         Yamaguchi 745-8585, Japan}

$^d${\it KEK Computing Research Center,
         High Energy Accelerator Research Organization,
         Ibaraki 305-0801, Japan} 

$^e${\it Graduate Institute for Advanced Studies, SOKENDAI,
         Ibaraki 305-0801, Japan} 

$^f${\it Department of Computer Science, Fukuyama University, Hiroshima 729-0292, Japan}

$^g${\it KEK Theory Center,
         High Energy Accelerator Research Organization,\\
         Ibaraki 305-0801, Japan} 

$^h${\it Research and Education Center for Natural Sciences,
         Keio University,\\
         Kanagawa 223-8521, Japan} 

$^i${\it Department of Physics,
         Shizuoka University,
         Shizuoka 422-8529, Japan} 

\end{center}

\vspace{2cm}

\clearpage
\thispagestyle{empty}
\addtocounter{page}{-1}

\begin{abstract}
  In our previous paper \cite{Ito:2020mys}, we found that the complex Langevin (CL) method works for QCD at finite density on the $16^3 \times 32$ lattice in the low-temperature high-density regime within the range $\mu / T = 1.6 - 9.6$ with $\mu$ and $T$ being the quark chemical potential and the temperature, which enabled us to see a clear trend towards the formation of the Fermi sphere. Here we investigate the validity of the CL method on the $24^3 \times 12$ lattice in the deconfined phase near the deconfinement phase transition. As before, we use four-flavor staggered fermions and judge the validity using the criterion based on the probability distribution of the drift term. The spatial extent is $L = (1.3 - 2.7 {\rm ~fm} )> \Lambda_{\rm LQCD}^{-1} \sim  1{\rm ~fm}$, in contrast to our previous study with $L < \Lambda_{\rm LQCD}^{-1}$. We find that the CL method works in a broad region up to $\mu / T = 4.8$, while it starts to fail as we approach the phase boundary due to the singular drift problem, which can be understood qualitatively by extending the Banks-Casher relation to the case at finite density.
\end{abstract}

\end{titlepage}

\tableofcontents

\renewcommand{\thefootnote}{\arabic{footnote}}
\setcounter{footnote}{0}

\section{Introduction}
\label{sec:introduction}
One of the long-standing issues in QCD is to elucidate the phase diagram
in the high-density region nonperturbatively from first principles.
It would enable us to understand quantitatively the experimental results of heavy-ion collisions,
the equation of state of neutron stars, and the color superconductivity predicted
by perturbation theory.
Although lattice simulation is expected to play a crucial role,
traditional Monte Carlo methods have been successful
only in the low-density region due to the sign problem.

The complex Langevin (CL) method~\cite{Klauder:1983sp,Parisi:1984cs} has been proposed as a potential approach to address the sign problem and has been successfully applied to QCD at finite density in specific regimes~\cite{Sexty:2013ica,Aarts:2014bwa,Fodor:2015doa,Sinclair:2015kva,Sinclair:2016nbg,Sinclair:2017zhn,Sinclair:2018rbk,Nagata:2018mkb,Ito:2018jpo,Tsutsui:2018jva,Tsutsui:2019gwn,Kogut:2019qmi,Sexty:2019vqx,Sinclair:2019ysx,Tsutsui:2019suq,Scherzer:2020kiu,Ito:2020mys} although the validity region is limited.
See Refs.~\cite{Berger:2019odf,Attanasio:2020spv,Nagata:2021ugx} for recent reviews.
It is based on the fictitious time evolution
of complexified dynamical variables by the Langevin equation,
which is an extension of the stochastic quantization \cite{Parisi:1980ys}
to the case with a complex action.
A unique equilibrium state of this stochastic process gives
the expectation value of physical observables, 
which agrees with that defined in the path integral under certain conditions.
Justification criteria of the CL method have been discussed in the continuous
Langevin time formulation~\cite{Aarts:2009uq,Aarts:2011ax,Nishimura:2015pba,Nagata:2015uga,Aarts:2017vrv,Scherzer:2018hid,Scherzer:2019lrh}
and the discretized Langevin time formulation~\cite{Nagata:2016vkn,Nagata:2018net}.
(See also Refs.~\cite{Salcedo:2016kyy,Cai:2020tgd} for related works.)
In particular, the latter formulation led to a practical validity condition
that the probability distribution of the drift term should fall off exponentially
or faster.
Since the drift term has to be calculated in the CL simulation anyway,
there is no additional cost associated with checking this validity condition.

In finite density QCD, the validity condition can be violated in the following two cases.
One is the case in which the complexified link variables deviate far from unitary matrices,
which is called the excursion problem \cite{Aarts:2009uq,Aarts:2011ax}.
The other is the case in which the Dirac operator for dynamical fermions acquires
near-zero eigenvalues with large probability \cite{Mollgaard:2013qra},
which is called the singular drift problem~\cite{Nishimura:2015pba}.
In these cases, either the gauge part or the fermionic part of the drift term
has a slow fall-off in the probability distribution.
It can also happen that one of the problems triggers the other.

The most important issue in practical applications of the CL method is
to identify
the validity region in the parameter space.
For that purpose, QCD with four-flavor staggered fermions provides a useful test bed
since it is numerically less costly than the Wilson fermions and
the phase structure has been investigated by various approaches~\cite{Fukugita:1990vu,Engels:1996ag,Fodor:2001au,DElia:2002tig,DElia:2004ani,Azcoiti:2004ri,Azcoiti:2005tv,deForcrand:2006ec,Fodor:2007vv,DElia:2007bkz,Li:2010qf,Endrodi:2018zda,Ito:2018jpo,Tsutsui:2018jva,Tsutsui:2019gwn,Tsutsui:2019suq,Ito:2020mys,Ohnishi:2015fhj}.
The deconfining phase transition is found to be of first-order at $\mu = 0$, and
it continues in the $\mu \neq 0$ region.
In the previous work \cite{Ito:2020mys}, we have shown that the CL method
works in finite density QCD on a $16^3 \times 32$ lattice at $\beta = 5.7$
with the bare quark mass $m = 0.01$ in the low-temperature high-density regime
within the range $\mu / T = 1.6 - 9.6$ with $\mu$ and $T$ being
the quark chemical potential and the temperature,
which enabled us to see a clear trend towards the formation of the Fermi sphere.

In this paper, we investigate the validity of the CL method
focusing on the region near the deconfining phase transition.
In Ref.~\cite{Fodor:2015doa}, the CL results in this region were compared
with those of the reweighting method.
The results obtained by the two methods
on the largest $16^3 \times 8$ lattice with $\mu / T = 0.96$
agree at $\beta \gtrsim 5.15$, whereas at smaller $\beta$, the CL simulation becomes unstable.
Interestingly, $\beta \sim 5.15$ is the point at which
the first-order deconfining phase transition occurs
according to the results of the reweighting method.

Motivated by this work, we perform CL simulations on a larger $24^3 \times 12$ lattice.
The spatial extent is $L = (1.3 - 2.7 {\rm ~fm} )> \Lambda_{\rm LQCD}^{-1} \sim  1{\rm ~fm}$,
in contrast to our previous study
with $L < \Lambda_{\rm LQCD}^{-1}$.
We explore the phase diagram within $0.6 \le \mu / T \le 4.8$,
which includes the high-density regime inaccessible by the reweighting method. 
We find that the CL simulation works in the deconfining phase even at $\mu / T = 4.8$
if the temperature is sufficiently high.
On the other hand, it starts to fail as one approaches the critical temperature
due to the singular drift problem.
We provide a qualitative interpretation of this behavior based on an
extension of the Banks-Casher relation to the case
at finite density \cite{Splittorff:2014zca,Nagata:2016alq}.
Some preliminary results of this work have been presented
in the proceedings article~\cite{Tsutsui:2018jva}.

This paper is organized as follows.
In section~\ref{sec:CL}, we explain
the CL method
and its application to QCD at finite density.
In section~\ref{sec:Results}, we show our results concerning the validity region
of the CL method based on the drift histogram.
Then we present our results for physical observables
obtained
within the validity region.
Section~\ref{sec:summary} is devoted to a summary and discussions.
In Appendix~\ref{sec:Banks-Casher}, we provide a brief
review on the Banks-Casher relation extended to finite density,
which explains why the CL method starts to fail near the
deconfining transition.

\section{The CL method for finite density QCD}\label{sec:CL}

In this section, we explain how one can apply
the CL method to finite density QCD.

\subsection{Finite density QCD with four-flavor staggered fermions}

Let us first define lattice QCD at finite density using four-flavor staggered fermions.
We introduce the link variables $U_{x,\nu} \in {\rm SU}(3) \, (\nu = 1,2,3,4)$ and the fermion fields $\bar{\chi}_x, \chi_x$, where $x = (x_1, x_2, x_3, x_4)$ represents the coordinates of each site.
We denote the spatial size of the lattice as $L_{\rm s}$ and the temporal size as $L_{\rm t}$.
The boundary conditions are taken to be periodic in all directions except that the
anti-periodic boundary condition is imposed on the fermion field in the temporal direction.
The temperature is then given by $T=(L_{\rm t})^{-1}$ in lattice units.

The lattice action is decomposed into the gauge part and the fermionic part as
\begin{align}
	S[U, \bar{\chi}, \chi] = S_\mathrm{g}[U] + S_\mathrm{f}[U, \bar{\chi}, \chi] \ .
\end{align}
For the gauge part $S_{\rm g}[U]$, we use the plaquette action defined by
\begin{align}
S_\mathrm{g}[U]
= -\frac{\beta}{6} \sum_{x}\sum_{\mu < \nu}
\mathrm{tr} \left( U_{x,\mu\nu} + U_{x,\mu\nu}^{-1} \right) \ ,
\quad
U_{x,\mu\nu} = U_{x,\mu} U_{x+\hat\mu,\nu} U_{x+\hat\nu,\mu}^{-1} U_{x,\nu}^{-1} \ ,
\end{align}
where $\beta$ is related to the gauge coupling constant $g$
through $\beta = 6 / g^2$, and $\hat\mu$ is a unit vector in the $\mu$-direction.
For the fermionic part $S_{\rm f}[U, \bar{\chi}, \chi]$, we use the staggered fermions defined by
\begin{align}
S_\mathrm{f}[U, \bar{\chi}, \chi]
&=
\sum_{x} \bar{\chi}_x M[U] \chi_x \ , \\
M[U] \, \chi_x
&=
\sum_{\nu = 1}^4 \frac{1}{2} \xi_{x,\nu} \left(
      e^{ \mu \delta_{\nu 4}} U_{x,\nu} \chi_{x+\hat{\nu}}
	- e^{-\mu \delta_{\nu 4}} U_{x-\hat{\nu},\nu}^{-1} \chi_{x-\hat{\nu}}
	+ m\chi_{x}
        \right) \ , \label{Mmat}
%
\end{align}
where $\xi_{x,\nu} = (-1)^{x_1 + \dots + x_{\nu-1}}$,
while $m$ and $\mu$ are the bare quark mass and the quark chemical potential, respectively,
in lattice units.
The partition function is defined by
\begin{align}
  Z = \int \prod_{x,\nu} \dd{U_{x,\nu}} \det M[U] \, e^{-S_\mathrm{g}[U]} \ ,
  \label{def-Z}
\end{align}
where the fermion determinant $\det M[U]$ is
complex
for $\mu \neq 0$, causing the sign problem.

\subsection{Application of the CL method}\label{sec:QCD CL}

When we apply the CL method to finite density QCD,
the link variables $U_{x,\nu} \in {\rm SU}(3)$ are replaced by the complexified link variables $\mathcal{U}_{x,\nu}\in {\rm SL}(3,\mathbb{C})$.
They evolve in fictitious time by the complex Langevin equation
\begin{align}
\mathcal{U}_{x,\nu}^{(\eta)}(t+\epsilon)=
\exp\left[ \im \left(-\epsilon \, v_{x,\nu}(\mathcal{U}^{(\eta)}(t))
+\sqrt{\epsilon} \, \eta_{x,\nu}(t)\right)\right]
\mathcal{U}_{x,\nu}^{(\eta)}(t) \ ,
\label{eq:cle}
\end{align}
where $t$ is the discretized Langevin time, $\epsilon$ is the stepsize, and $\eta_{x,\nu}(t)$ is a $3\times3$ traceless Hermitian matrix generated with the Gaussian distribution $\exp(- \mathrm{tr} \{ \eta_{x,\nu}^{2}(t) \} /4)$.
The so-called drift term $v_{x,\nu}(\mathcal{U})$ is defined through the effective action as
\begin{align}
v_{x,\nu}(\mathcal{U})
&= \sum_{a=1}^8  \lambda_a
\left. \dv{\alpha}
S_\mathrm{eff}[e^{\im \alpha \lambda_a}\mathcal{U}_{x ,\nu}]
\right|_{\alpha=0} \ ,
\label{drift term} \\
S_\mathrm{eff}[\mathcal{U}] &= S_\mathrm{g}[\mathcal{U}] - \log \det M(\mathcal{U}) \ ,
\label{effective action}
\end{align}
where $\lambda_a \ (a=1,\cdots,8)$ are the SU(3) generators normalized by $\mathrm{tr}(\lambda_a \lambda_b) = \delta_{ab}$.
For later convenience, we decompose the drift term into the gauge and fermionic parts as
\begin{align}
v_{x,\nu}^\text{(g)} 
&= \sum_{a=1}^8 \lambda_a
\left.\dv{\alpha} S_\mathrm{g}[e^{\im \alpha \lambda_a}\mathcal{U}_{x,\nu}]\right|_{\alpha=0} \ , \\
v_{x,\nu}^\text{(f)} 
&= \sum_{a=1}^8 \lambda_a
\left.\dv{\alpha} \left( - \log \det M(e^{\im \alpha \lambda_a}\mathcal{U}_{x,\nu}) \right)\right|_{\alpha=0} \ .
\label{def-drift-fermi}
\end{align}
For the computation of the fermionic part of the drift term, we used a standard conjugate gradient (CG) solver without any preconditioning or deflation techniques.

The validity of the CL method implies that
the expectation value of
a holomorphic observable $O(\mathcal{U})$
with respect to the partition function \eqref{def-Z}
can be evaluated
as
\begin{align}
\expval{O(\mathcal{U})} =
\lim_{t\rightarrow\infty}
\expval{O(\mathcal{U}^{(\eta)}(t))}_{\eta} \ .
\label{eq:expect_value}
\end{align}
Here $\expval{ \ \cdot \ }_{\eta}$ is an average over the Gaussian noise $\eta$,
which is replaced in practice by the time average for one sequence of $\eta_{x,\nu}(t)$
assuming ergodicity.
A proof of eq.~\eqref{eq:expect_value} was discussed in Refs.~\cite{Aarts:2009uq,Aarts:2011ax} through a direct analysis of the boundary term, providing the first concrete framework for validating the CLM. Following these studies, Ref.~\cite{Nagata:2016vkn} refined the discussion and provided a practical criterion for the validity of the CL method that the probability distribution of the drift term should fall off exponentially or faster at large magnitudes.
Consistency of this criterion based on the drift term with that based on the boundary term \cite{Aarts:2009uq,Aarts:2011ax,PhysRevD.109.014509} was confirmed in Ref.~\cite{Scherzer:2018hid}.
In practice, we define the maximum magnitude of the drift term as
\begin{align}
  v_{\rm g}  = \sqrt{\frac{1}{3} \max_{x, \nu}\mathrm{tr}
    \Big(v_{x,\nu}^{{\rm (g)} \dagger} v_{x,\nu}^{{\rm (g)}} \Big)  } \ , \quad \quad
      v_{\rm f}  = \sqrt{\frac{1}{3} \max_{x, \nu}\mathrm{tr}
    \Big(v_{x,\nu}^{{\rm (f)} \dagger} v_{x,\nu}^{{\rm (f)}} \Big) } \ ,
\end{align}
and check whether the probability distributions $p(v_{\rm g})$, $p(v_{\rm f})$
fall off exponentially or faster.

As we mentioned in section \ref{sec:introduction},
there are two cases
in which the criterion with the drift term is not satisfied.
One of them \cite{Aarts:2009uq,Aarts:2011ax} is referred to as the excursion problem,
which occurs when the complexified link variables make a long excursion
away from the ${\rm SU}(3)$ manifold.
As a measure of the distance from the ${\rm SU}(3)$ manifold, we define the unitarity norm
\begin{equation}
{\cal N}=\frac{1}{12 V}\sum_{x,\nu}
\mathrm{tr}(\mathcal{U}_{x,\nu}^{\dagger} \mathcal{U}_{x,\nu}-{\bf 1})\ ,
\label{def-unitarity-norm}
\end{equation}
where $V=(L_{\rm s})^3 L_{\rm t}$ is the total number of lattice sites.
This problem can be reduced in many cases by
the gauge cooling \cite{Seiler:2012wz}, which amounts to making
a complexified gauge transformation
\begin{equation}
\mathcal{U}_{x,\nu} \rightarrow
g_{x} \, \mathcal{U}_{x,\nu} \,  g_{x+\hat{\nu}}^{-1} \ , \quad \quad
g_{x} \in {\rm SL}(3,\mathbb{C})
\end{equation}
in such a way that the unitarity norm \eqref{def-unitarity-norm}
is minimized after each step
of updating $\mathcal{U}_{x,\nu}$ by the complex Langevin equation~\eqref{eq:cle}.
(See Refs.~\cite{Nagata:2015uga,Nagata:2016vkn} for justification.)
To mitigate the excursion problem, we applied this gauge cooling technique.
The other case \cite{Mollgaard:2013qra,Nishimura:2015pba} is referred to as
the singular drift problem,
which is caused by the near-zero eigenvalues of $M$ due to the
fermionic part \eqref{def-drift-fermi} of the drift term involving $M^{-1}$.

The observables we investigate in this work are the chiral condensate,
the Polyakov loop and the quark number.
The chiral condensate is defined by 
\begin{eqnarray}
 \Sigma 
 &=& \frac{1}{ V }
     \frac{\partial}{\partial m}\log Z
 = \frac{1}{ V } \left\langle \mathrm{Tr} \, M^{-1} \right\rangle \ ,
 \label{eq:chiral-condensate}
\end{eqnarray}
where the trace ${\rm Tr}$ is taken for both color and spacetime indices.
The Polyakov loop is defined by
\begin{equation}
  P=\frac{1}{3(L_{\mathrm{s}})^{3}} \sum_{\vec{x}}
  \left\langle
  \mathrm{tr} 
 \left( \prod_{x_4=1}^{L_{\rm t}} U_{(\vec{x},x_4),4} 
 \right)
 \right\rangle
 \ , \label{eq:polyakov}
\end{equation}
where $\vec{x}=(x_{1},x_{2},x_{3})$ represents the coordinates in spatial directions.
The quark number is defined by 
\begin{eqnarray}
 N_{\rm q} 
 = \frac{1}{L_{\rm t}} \frac{\partial}{\partial \mu}\log Z
 = \frac{1}{L_{\rm t}}
 \left\langle 
 \sum_{x}\frac{\xi_{x,4}}{2}
 {\rm Tr}
 \left(e^{\mu}M_{x+\hat{4},x}^{-1}U_{x,4}
 +e^{-\mu}M_{x-\hat{4},x}^{-1}U_{x-\hat{4},4}^{-1}\right)\right\rangle \ .
 \label{eq:quark-number}
\end{eqnarray}

\section{Results}
\label{sec:Results}

In this section, we show our results on a $24^3 \times 12$ lattice at $\beta = 5.2 - 5.5$
for $\mu / T = 0.6 - 4.8$
with fixed bare quark mass $m = 0.01$.
The Langevin stepsize is set to $\epsilon = 5 \times 10^{-5}$ initially and changed adaptively when the drift term exceeds a threshold~\cite{Aarts:2009dg}.
Here the stepsize is replaced by $\epsilon v_\text{th}/v $
if $v > v_\text{th}$ is satisfied, where $v = \max_{x,\nu} |v_{x,\nu}|$ and $v_\text{th} = 0.01$.
We accumulate $12000-32000$ configurations for the measurement of observables during the Langevin time $t = 2-11$ after thermalization. For details on the thermalization criterion, configuration sampling interval, and autocorrelation analysis, see Appendix~\ref{app:thermalization}.
The lattice spacing $a$ for each $\beta$ is determined as given in Table \ref{tab:lattice-setup}
by interpolating the results
at $\beta \leq 5.4$~\cite{Fodor:2015doa} and $\beta = 5.7$~\cite{Ito:2020mys}.
In Table \ref{tab:lattice-setup}, we also list
the spatial extent of the lattice $L_{\rm s} a= 24 a$
and the temperature $T=1/(a L_{\rm t})=1/(12 a)$
for our setup at each $\beta$.
We note that the physical pion mass for our bare quark mass $m=0.01$ is approximately 519 MeV at $\beta = 5.2$ and 742 MeV at $\beta = 5.4$~\cite{Fodor:2015doa}.

\begin{table}[t]
\centering
\begin{tabular}{cccc}
\hline
$\beta$ & $a$ [fm] & $L_{\rm s} a$ [fm]  & $T$ [MeV] \\ \hline
5.2 & 0.112 & 2.70  & 146 \\
5.3 & 0.084 & 2.01  & 196 \\
5.4 & 0.065 & 1.56  & 252 \\
5.5 & 0.053 & 1.28  & 308 \\
\hline
\end{tabular}
\caption{The lattice spacing $a$,
  the spatial extent of the lattice $L_{\rm s} a= 24 a$
  and the temperature $T=1/(a L_{\rm t})=1/(12 a)$
  are given for our setup at each $\beta$.}
\label{tab:lattice-setup}
\end{table}

\begin{figure}[t]
	\centering
	\includegraphics[width=7.5cm]{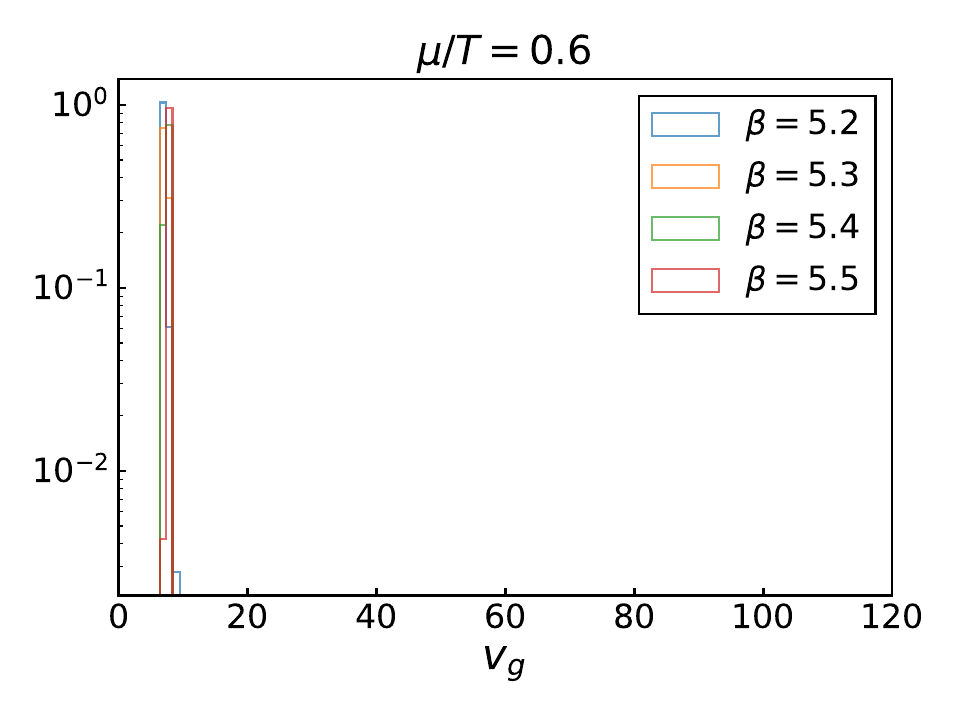}
	\includegraphics[width=7.5cm]{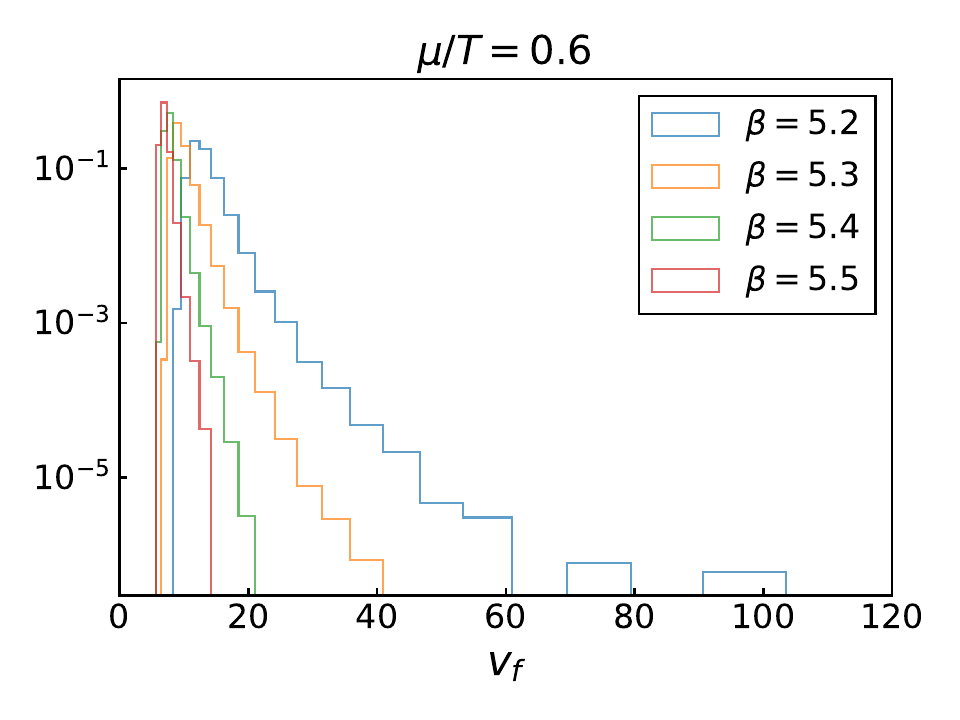}
	\caption{The drift histogram
          for the gauge part (Left) and the fermionic part (Right)
is shown for $\mu / T = 0.6$ with $5.2 \le \beta \le 5.5$.}
	\label{fig:drift_mu0.1}
\end{figure}

\subsection{Validity region in the $(\mu, T)$-plane}
\label{sec:validity}

Let us first discuss the validity of the CL method based on the drift histogram.
In Fig.~\ref{fig:drift_mu0.1}, we show the drift histogram
for the gauge and fermionic parts for $\mu / T = 0.6$.
For the gauge part, we observe a fast fall-off.
For the fermionic part, on the other hand, we observe a long power-law tail
at $\beta = 5.2$.
Similar results are obtained for other values of $\mu / T$.
Thus the singular drift problem is the main cause of the failure of the CL method at small $\beta$. (The unitarity norm will be discussed later.)

\begin{table}[t]
	\centering
	\begin{tabular}{ccccccc}
		\hline
		$\mu/T$ & $\beta$ & \phantom{..} $p$ &\phantom{..} $\alpha$ & $\chi^2/\text{d.o.f.}(p)$ & $\chi^2/\text{d.o.f.}(\alpha)$ & CG method \\ \hline
		0.6 & 5.2 & $7.4\pm 0.3$  & $5.8\pm 0.3$ & $0.093-0.14$ & $0.14-0.22$ &        \\
		0.6 & 5.3 & $9.6\pm 0.3$  & $2.6\pm 0.2$ & $0.18-0.25$ & $0.29-0.51$ &        \\
		0.6 & 5.4 & $13.3\pm 0.9$ & $1.14\pm 0.03$ & $0.053-0.48$ & $0.052-0.35$ &       \\
		0.6 & 5.5 & $15.8\pm 0.1$ & $0.6\pm 0.1$ & $0.033-0.035$ & $0.38-1.5$ &       \\
		1.2 & 5.3 & $3.21\pm 0.02$  & $23.1\pm 1.5$ & $0.013-0.026$ & $0.031-0.043$ &        \\
		1.2 & 5.4 & $4.1\pm 0.3$  & $6.6\pm 1.0$ & $0.25-0.26$ & $0.39-0.48$ &         \\
		1.2 & 5.5 & $2.09\pm 0.05$ & $26.5\pm 1.5$ & $0.062-0.068$ & $0.080-0.095$ &       \\
		2.4 & 5.2 & - & - & - & - & not converged  \\
		2.4 & 5.3 & - & - & - & - & not converged   \\
		2.4 & 5.4 & $9.1\pm 0.2$ & $2.9\pm 0.2$ & $0.0062-0.027$ & $0.011-0.11$ &        \\
		2.4 & 5.5 & $14.3\pm 0.2$ & $0.8\pm 0.1$ & $0.24-0.34$ & $0.66-1.1$ &       \\
		3.6 & 5.2 & -  & - & - & - & not converged  \\
		3.6 & 5.3 & - & - & - & - & not converged  \\
		3.6 & 5.4 & $10.9\pm 0.2$ & $2.3\pm 0.3$ & $0.061-0.078$ & $0.00044-0.66$ &        \\
		3.6 & 5.5 & $4.7\pm 0.6$ & $7.9\pm 1.6$ & $0.09-0.11$ & $0.12-0.15$ &        \\
		4.8 & 5.2 & -  & - & - & - & not converged \\
		4.8 & 5.3 & -  & - & - & - & not converged  \\
		4.8 & 5.4 & - & - & - & - & not converged \\
		4.8 & 5.5 & $9.3\pm 0.8$ & $4.0\pm 0.9$ & $0.19-0.39$ & $0.19-1.1$ &        \\
		\hline
	\end{tabular}
	\caption{The result of fitting the histogram of the fermionic drift
          to a power law $x^{-p}$ and an exponential function $e^{-x/\alpha}$. The values of
          $\chi^2$/d.o.f. listed in the table represent the range obtained by varying the fit range of the data. 
	  The right-most column indicates the case in which the CG solver is not converged
          and hence the thermalization is not achieved.} 
	\label{tab:fitting}
\end{table}

In order to distinguish between the exponential fall-off and the power-law fall-off
in the drift histogram,
we fit the tail of the histogram to a power law $v^{-p}$ and an exponential function
$e^{-v/\alpha}$.
The bin size is chosen to be uniform in the log scale
in order to accumulate sufficient statistics in the tail of the distribution.
The error for each bin is estimated correctly by the jackknife method
although the correlation between each bin is ignored in fitting for simplicity.
The fitting range is chosen
in such a way that 
the values of the fitting parameters are insensitive
to the choice.
As a result, the number of bins used for fitting is chosen as
$n_{\rm fit}=10,11,12$ for $\beta=5.4$, $5.5$ and $\mu/T=1.2$, $n_{\rm fit}=4,5$ for $\beta=5.5$ and $\mu/T=0.6$, and $n_{\rm fit}=4,5,6$ otherwise\footnote{In fitting the data for the $n_{\rm fit}$ bins
corresponding to large values of the drift term,
we exclude the isolated bins, which appear when
the configuration approaches the singularity of the drift term accidentally
depending on the initial configuration.
Such events are very rare and they do not affect the validity criterion of the CL method.}.

In Table~\ref{tab:fitting} we present the results of the fitting
for 19 different parameter sets used in our CL simulation.
The errors in the parameters $p$ and $\alpha$ include both the fitting errors and the uncertainty arising from variations in the fitting range.
The last column indicates the case in which the CG method used to solve the linear equation associated with the fermion matrix does not converge and hence the thermalization is not achieved.

\begin{figure}[H]
	\centering
	\includegraphics[width=6cm]{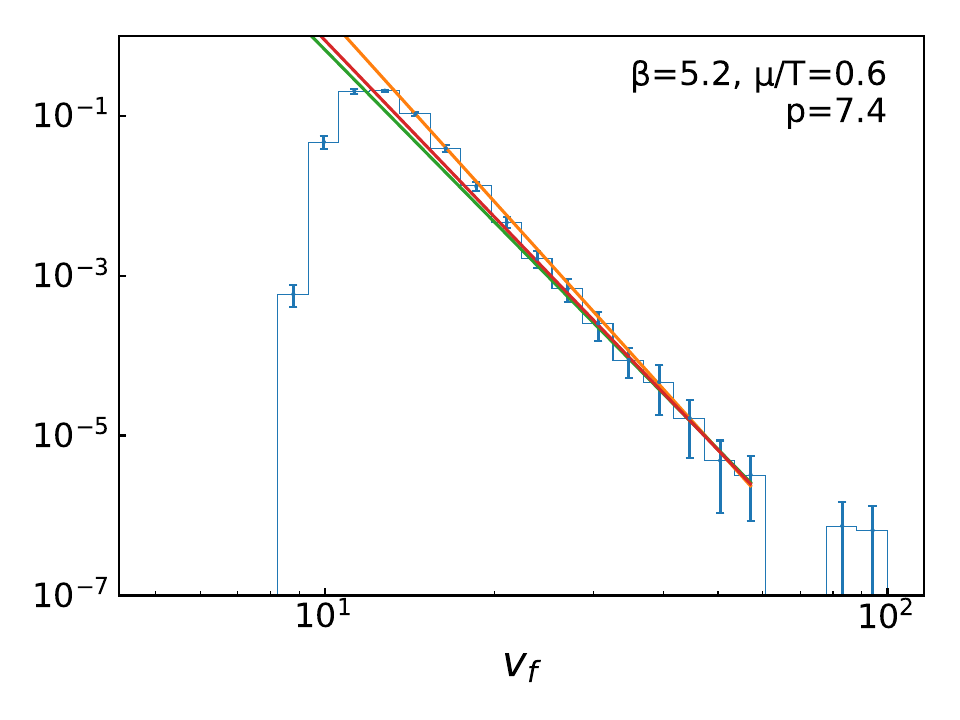}
	\includegraphics[width=6cm]{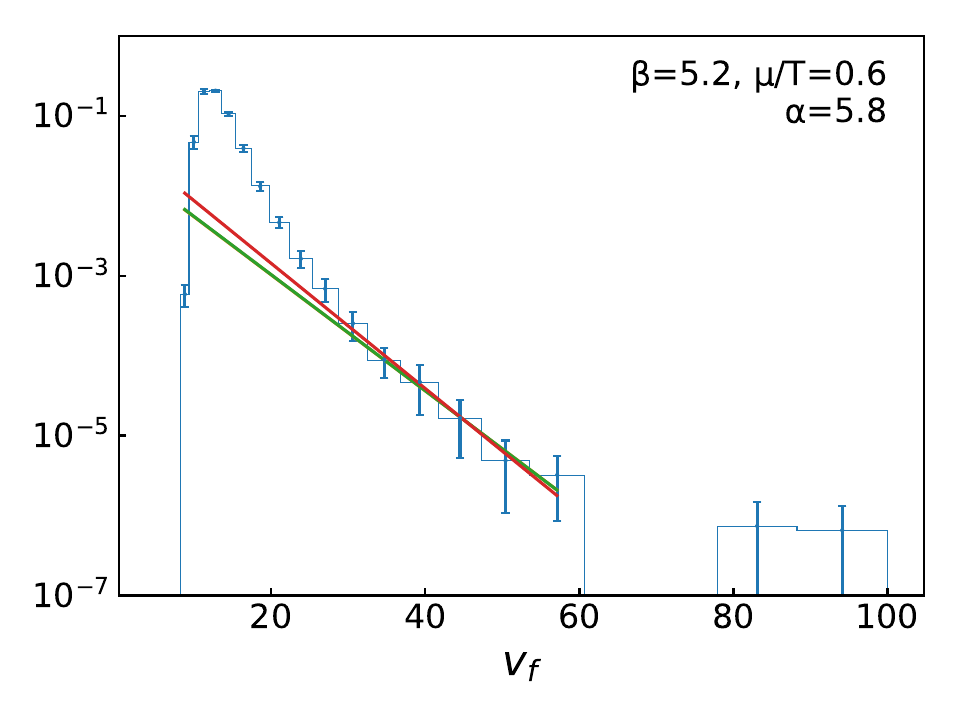}
	\includegraphics[width=6cm]{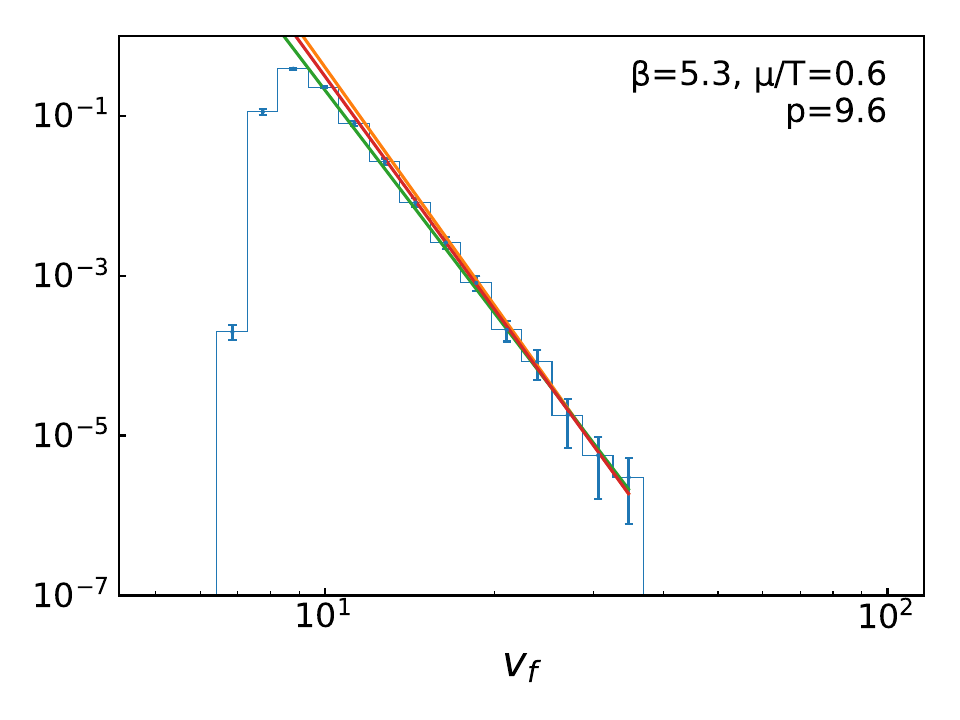}
	\includegraphics[width=6cm]{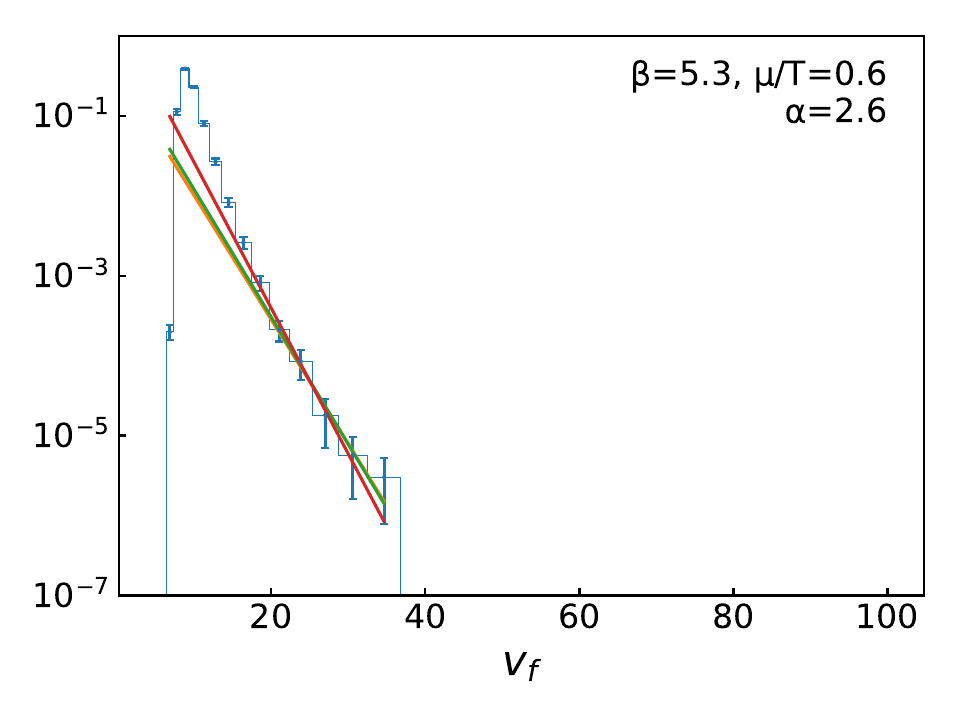}
	\includegraphics[width=6cm]{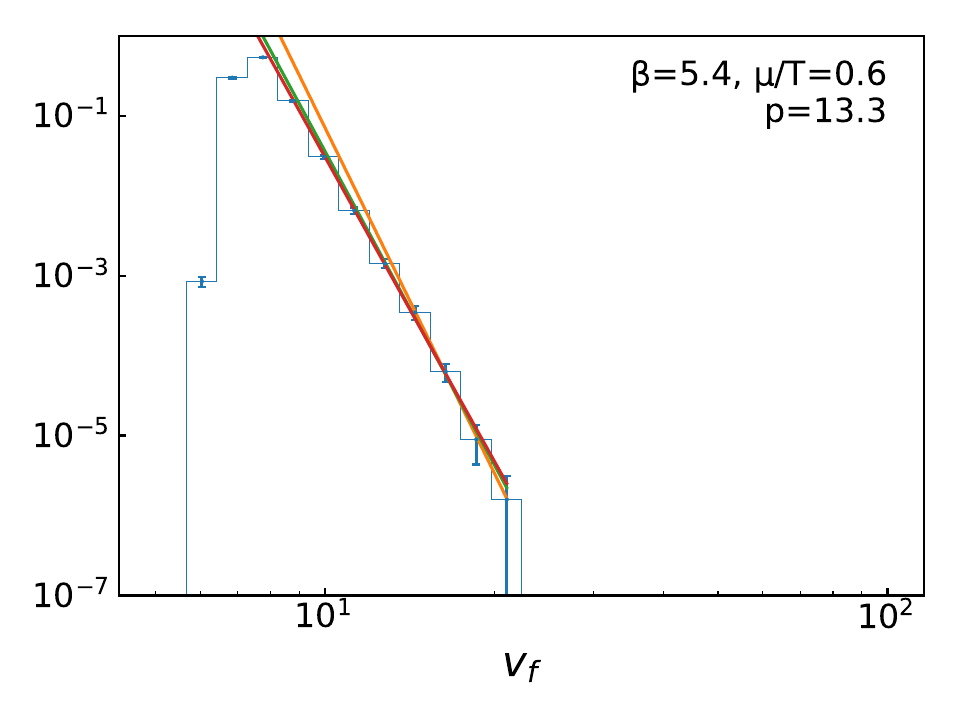}
	\includegraphics[width=6cm]{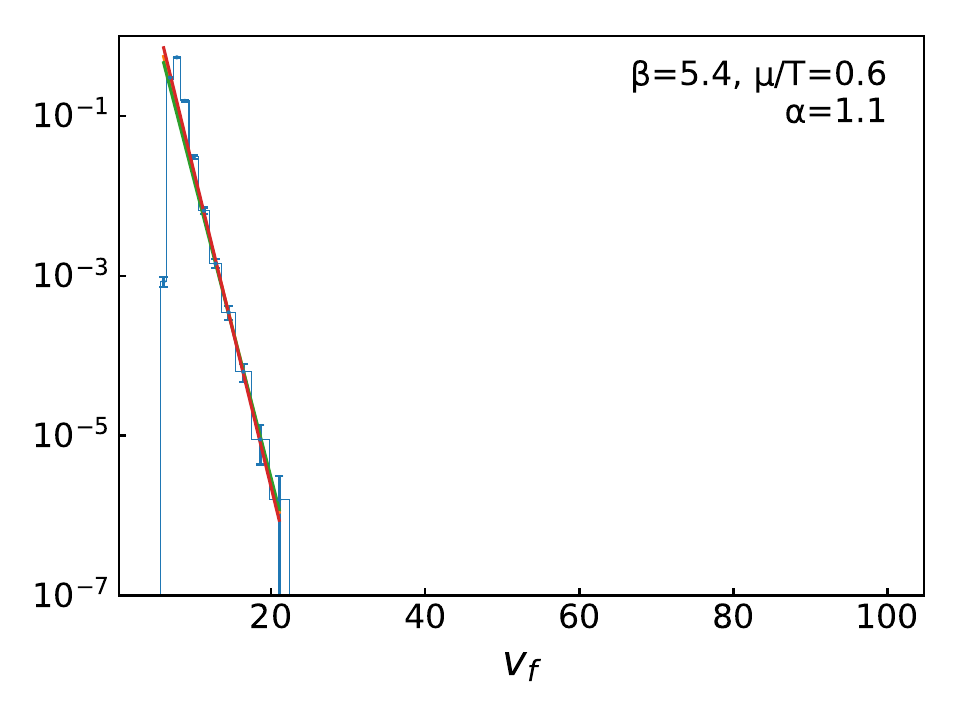}
	\includegraphics[width=6cm]{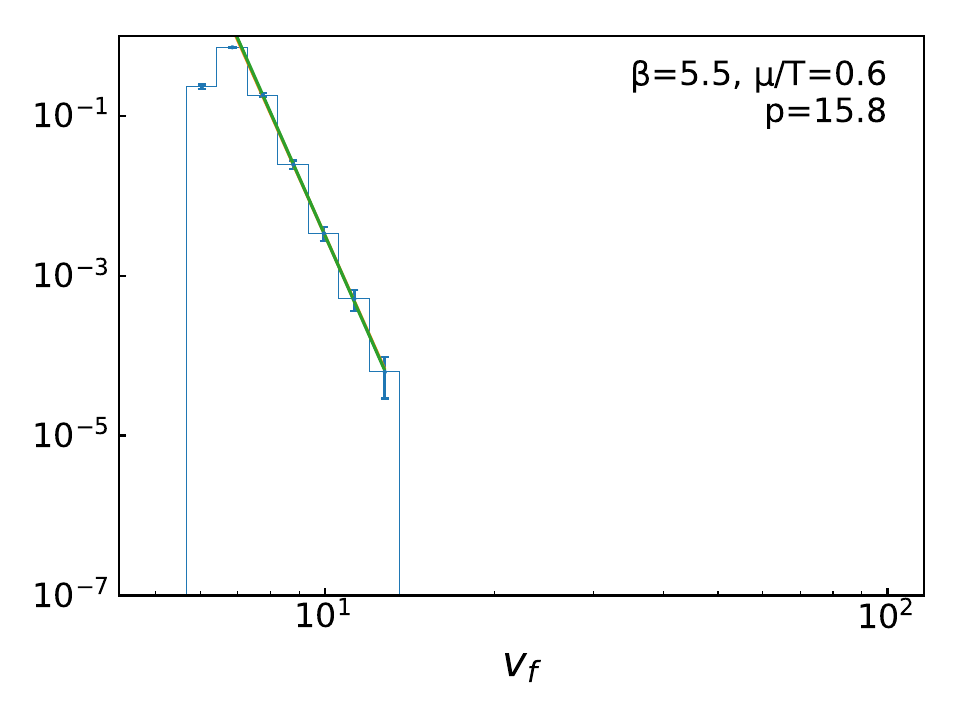}
	\includegraphics[width=6cm]{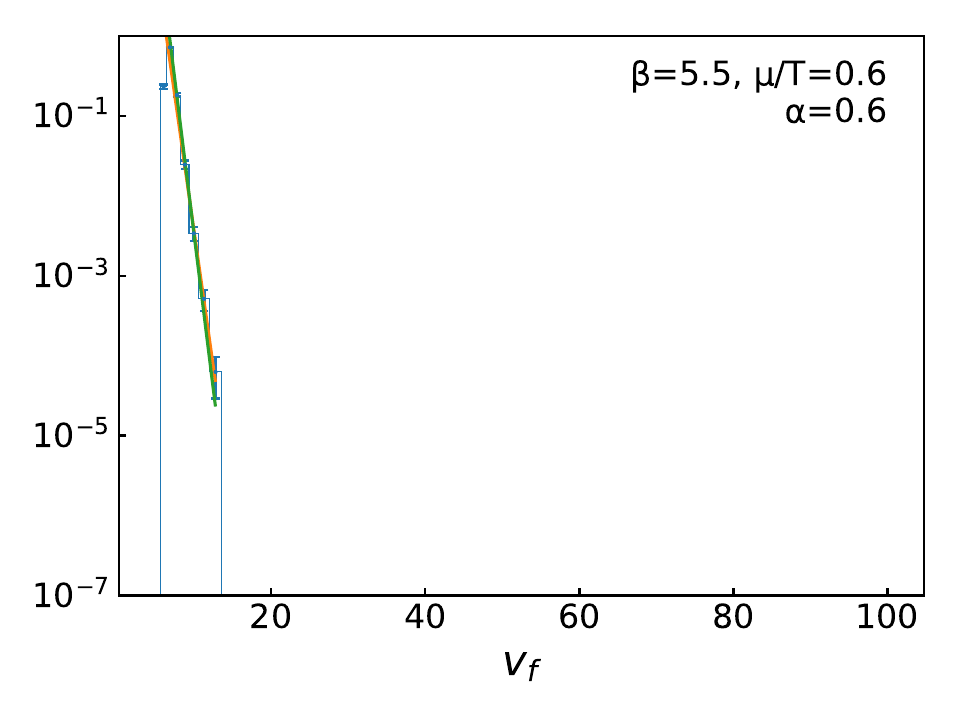}
	\caption{Histogram of the fermionic drift is shown
          in the log-log scale (Left) and the semi-log scale (Right)
          for $\mu/T=0.6$ at $5.2 \le \beta \le 5.5$.
          The solid lines represent the fits to a power law $x^{-p}$ (Left)
          and an exponential function $e^{-x/\alpha}$ (Right).
          Different colors indicate the results with a different fitting range.}
	\label{drift_fit_mu0.05}
\end{figure}

In Fig.~\ref{drift_fit_mu0.05}
we show the histogram of the fermionic drift
for $\mu / T = 0.6$ at $5.2 \le \beta \le 5.5$
together with the fits to a power law (Left) and an exponential function (Right).
By comparing the quality of the fits,
we find that a power-law fall-off is more likely for $\beta = 5.2$,
whereas an exponential fall-off is more likely for $\beta = 5.3$, $5.4$ and $5.5$,
which seems to be supported also by the values for the fitting parameters
$p$ and $\alpha$ obtained by the fits.
Thus we conclude that the CL method is valid for $\mu / T = 0.6$ 
at $\beta = 5.3$, $5.4$ and $5.5$ but not at $\beta = 5.2$.

\begin{figure}[t]
	\centering
	\includegraphics[width=12cm]{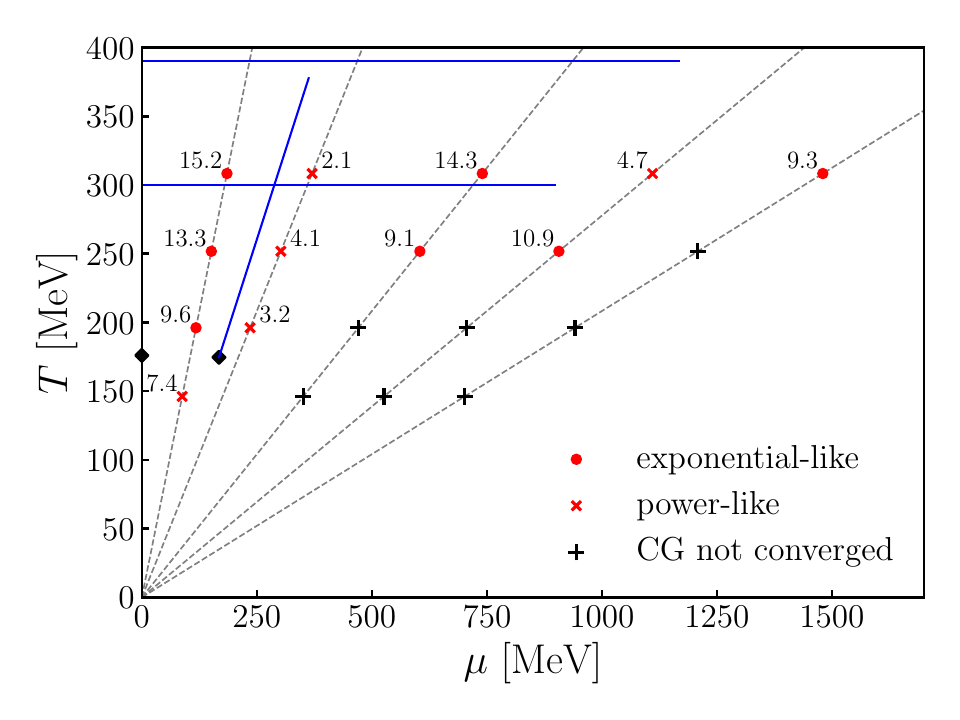}
	\caption{The validity region of the CL method is shown
          in the $(\mu,T)$-plane, where the dashed lines correspond
          to $\mu / T = 0.6, 1.2, 2.4, 3.6$ and $4.8$.
          The circles (crosses) represent the parameter points on which
          the distribution of the drift term can (not) be fitted by an exponential function.
          The numbers close to these symbols represent the power
          obtained by fitting the tail of the distribution to a power law.
        The plus symbols indicate that the CG method does not converge at these points.
        The diamonds represent the critical temperature~\cite{Engels:1996ag,Fodor:2015doa}
        obtained at two different values of $\mu$ assuming that the string tension
        is given by $\sqrt{\sigma} = 440$ MeV.
        The diagonal and horizontal solid blue lines correspond to the parameter sets
        used in Ref.~\cite{Fodor:2015doa} and Ref.~\cite{Sexty:2019vqx}, respectively.}
	\label{fig:summary of validity at m=0.01}
\end{figure}

Figure \ref{fig:summary of validity at m=0.01} summarizes the validity region
of the CL method in the $\mu$-$T$ plane.
We find that the validity region extends broadly in the deconfined phase.
As the temperature becomes lower, however, the CL method starts to fail and
eventually the thermal equilibrium cannot be achieved because the CG method does not converge.
The solid blue lines indicate the parameter sets used
in the previous studies~\cite{Fodor:2015doa,Sexty:2019vqx}.
While the validity condition was not discussed there,
the observation in Ref.~\cite{Fodor:2015doa}
that the simulations start to fail as one approaches the critical temperature
is consistent with our finding.
The phase diagram was investigated by the CL method
also with different lattice actions~\cite{Kogut:2019qmi, Sexty:2019vqx, Scherzer:2020kiu}.

Here we point out that
the appearance of a large fermionic drift
can be understood qualitatively by generalizing the Banks-Casher relation \cite{Banks:1979yr}
to the case at
finite density \cite{Splittorff:2014zca,Nagata:2016alq}.
(See Appendix \ref{sec:Banks-Casher} for a brief review.)
This relation implies that, in a phase with nonzero chiral condensate, the eigenvalues
of the Dirac operator in the CL simulation should distribute around zero
as one takes the infinite-volume and chiral limits.

\clearpage

\begin{figure}[H]
	\centering
	\includegraphics[width=7.5cm]{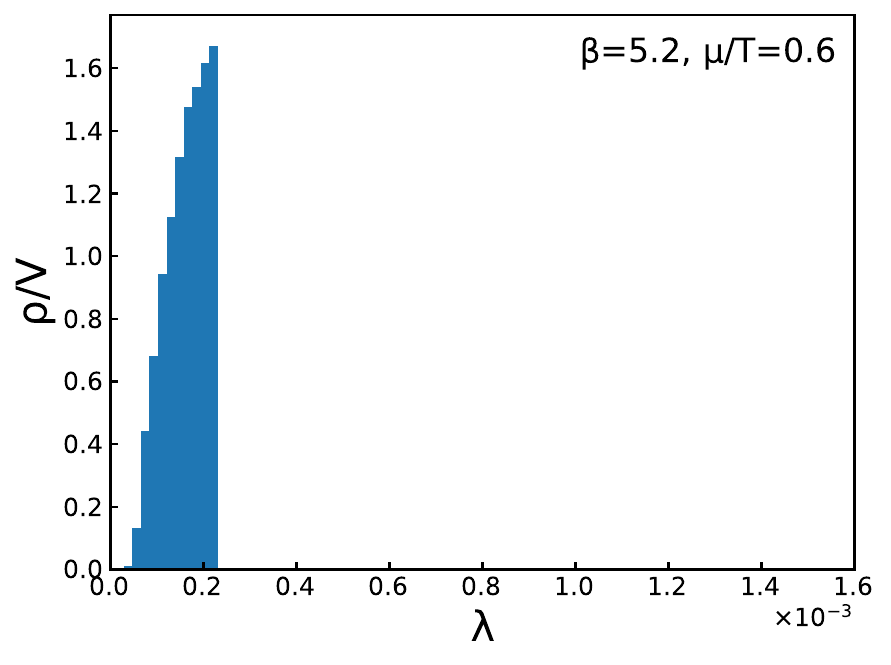}
	\includegraphics[width=7.5cm]{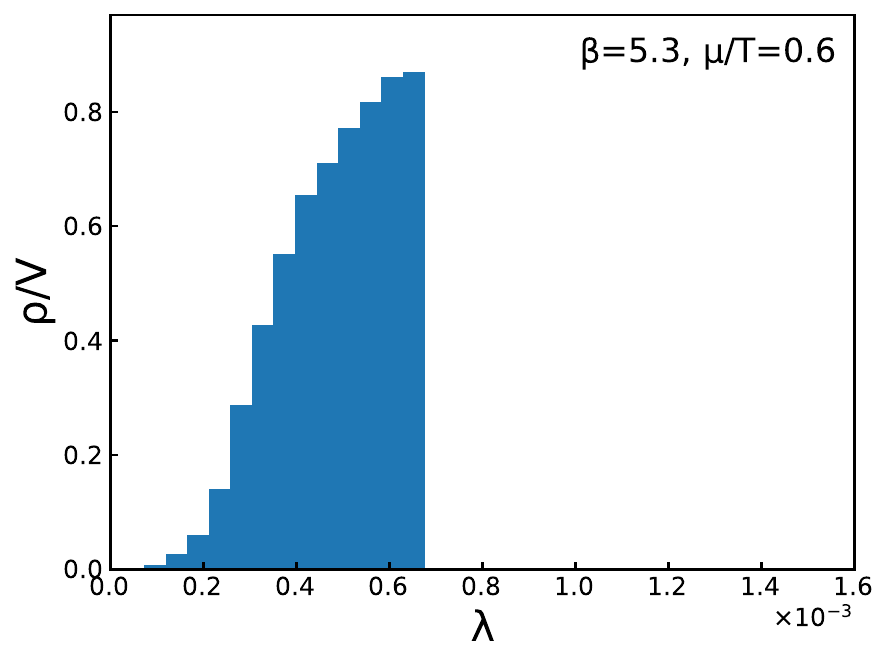}
	\includegraphics[width=7.5cm]{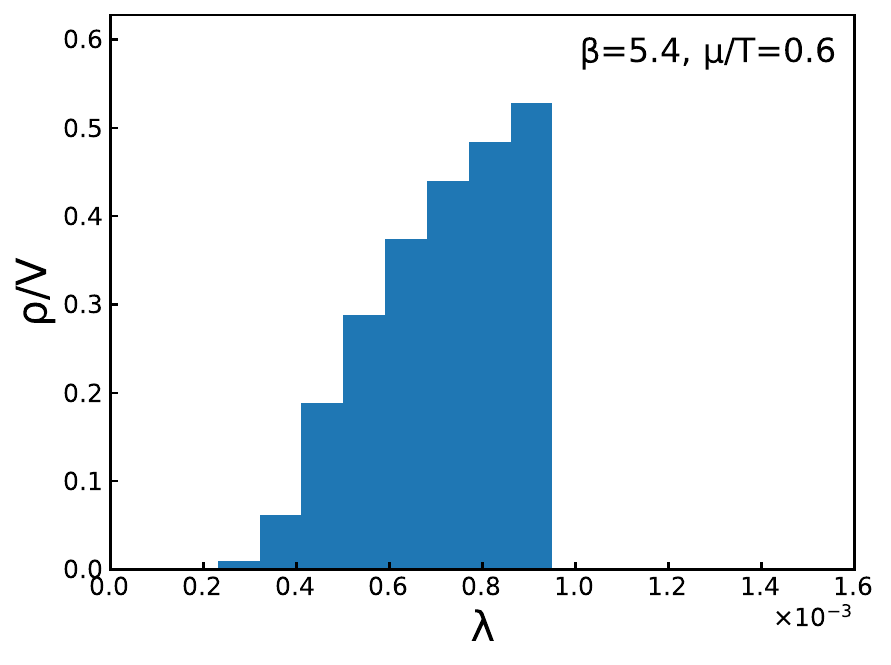}
	\includegraphics[width=7.5cm]{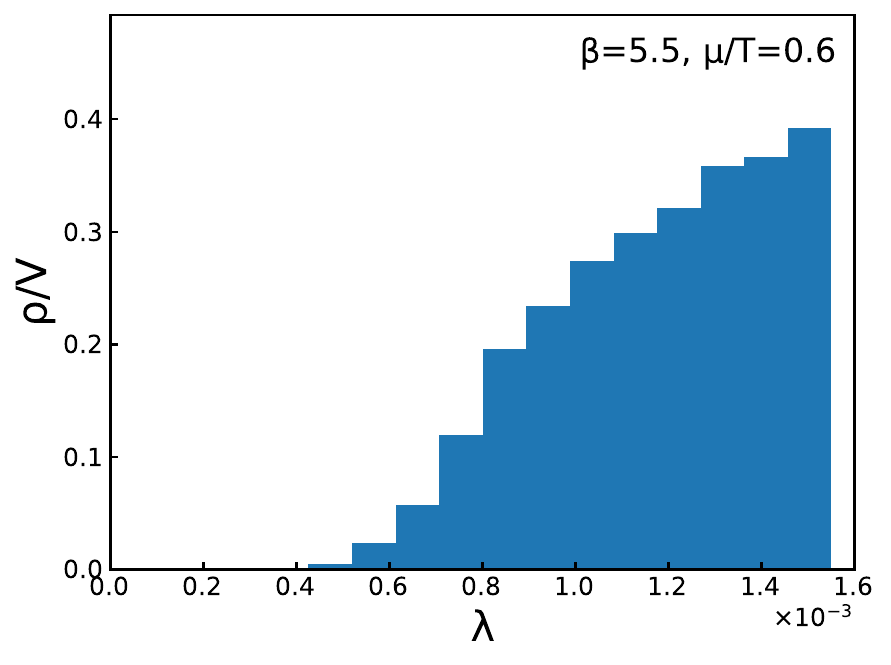}
	\caption{The eigenvalue distribution of $M^\dagger M$, 
         where $M$ is the Dirac operator defined in \eqref{Mmat},
         is plotted for $\mu/T = 0.6$ at $\beta=5.2, 5.3, 5.4$ and $5.5$.}
	\label{fig:eigvals}
\end{figure}

In Fig.~\ref{fig:eigvals}, we show the distribution of
the low-lying eigenvalues of $M^\dagger M$ for $\mu/T=0.6$
obtained by the standard implicitly restarted Lanczos algorithm\footnote{We have used
a code in Bridge++~\cite{Ueda:2014rya} for this calculation.},
where $M$ is the Dirac operator defined in \eqref{Mmat}.
While our simulation is performed on a finite $24^3 \times 12$ lattice
and with nonzero quark mass $m=0.01$, it is conceivable that
the Dirac operator acquires near-zero eigenvalues, which cause a large fermionic drift
as we approach the
boundary of the deconfined phase
by lowering $\beta$.

\begin{figure}[H]
	\centering
	\includegraphics[width=9cm]{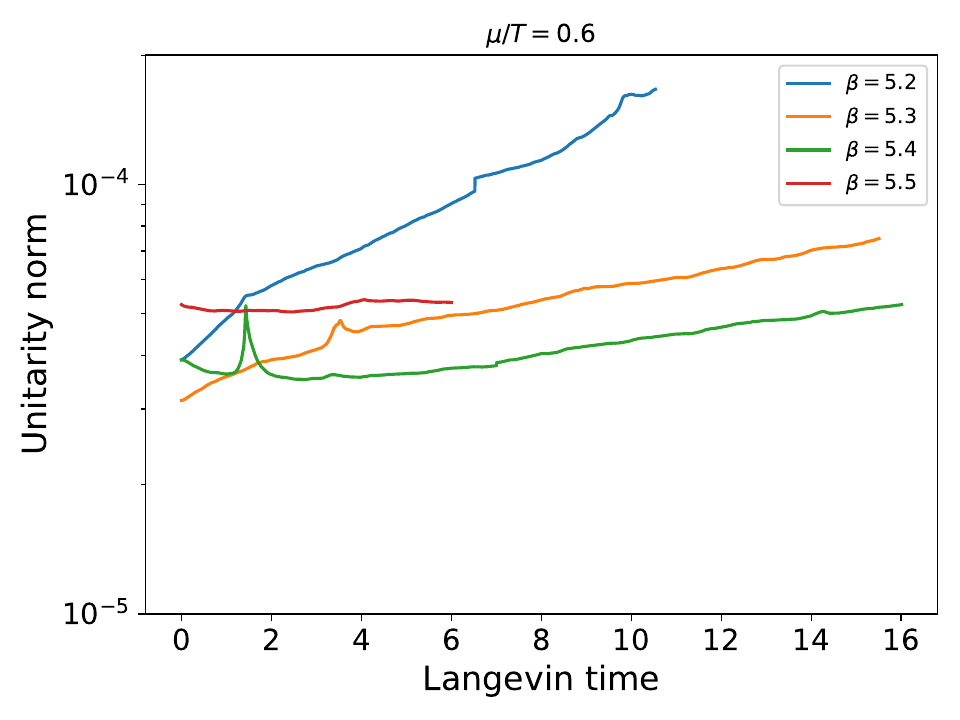}
	\caption{The Langevin-time history of the unitarity norm \eqref{def-unitarity-norm}
          is plotted for $\mu / T = 0.6$.}
	\label{fig:unitarity norm}
\end{figure}

Let us also discuss the Langevin-time history of the unitarity norm \eqref{def-unitarity-norm},
which we monitor as an indicator of the excursion problem.
In Fig.~\ref{fig:unitarity norm}, we show the history of the unitarity norm for $\mu/T=0.6$.
While the unitarity norm at $\beta=5.2$ shows rapid growth,
we do not observe a power-law tail in the histogram of the gauge part of the drift term
as shown in Fig.~\ref{fig:drift_mu0.1} (Left).
A similar phenomenon is observed in CL simulations of
2D U(1) gauge theory with a $\theta$ term~\cite{Hirasawa:2020bnl}.
In fact, it is not obvious how one can detect the excursion problem
from the growth of the unitarity norm.
The unitarity norm at $\beta > 5.2$ also increases with time but
more slowly 
than at $\beta = 5.2$.
From the above observations, we conclude that
the excursion problem does not seem to occur in our CL simulations
for $\mu/T=0.6$ at $5.2 \leq \beta \leq 5.5$,
at least within the current statistics.

\subsection{Physical observables}
\label{sec:observables}

Let us then discuss our results for physical observables,
the chiral condensate \eqref{eq:chiral-condensate},
the Polyakov loop \eqref{eq:polyakov} and
the quark number \eqref{eq:quark-number},
obtained
within the validity region of the CL method.
The trace ${\rm Tr}$ in Eqs.~\eqref{eq:chiral-condensate} and \eqref{eq:quark-number}
is evaluated by the standard noisy estimator using 20 noise vectors.
We note that the number of effectively independent configurations is limited, particularly for $\beta = 5.3$ and $5.4$, due to long autocorrelation times. As such, the results presented here should be regarded as indicative. A detailed discussion on the autocorrelation analysis is found in Appendix~\ref{app:thermalization}.

In Fig.~\ref{fig:order parameters at m=0.01}, we plot the chiral condensate (Left)
and the Polyakov loop (Right) against $\mu/T$.
We find that the chiral condensate tends to grow and the Polyakov loop tends to
decrease as $\beta$ decreases.
This is understandable since lowering $\beta$ implies approaching
the chiral-symmetry-broken confined phase.
The CL method starts to fail at some point due to
the near-zero eigenvalues of the Dirac operator
as we have seen in Fig.~\ref{fig:eigvals}.

\begin{figure}[H]
	\centering
	\includegraphics[width=8.0cm]{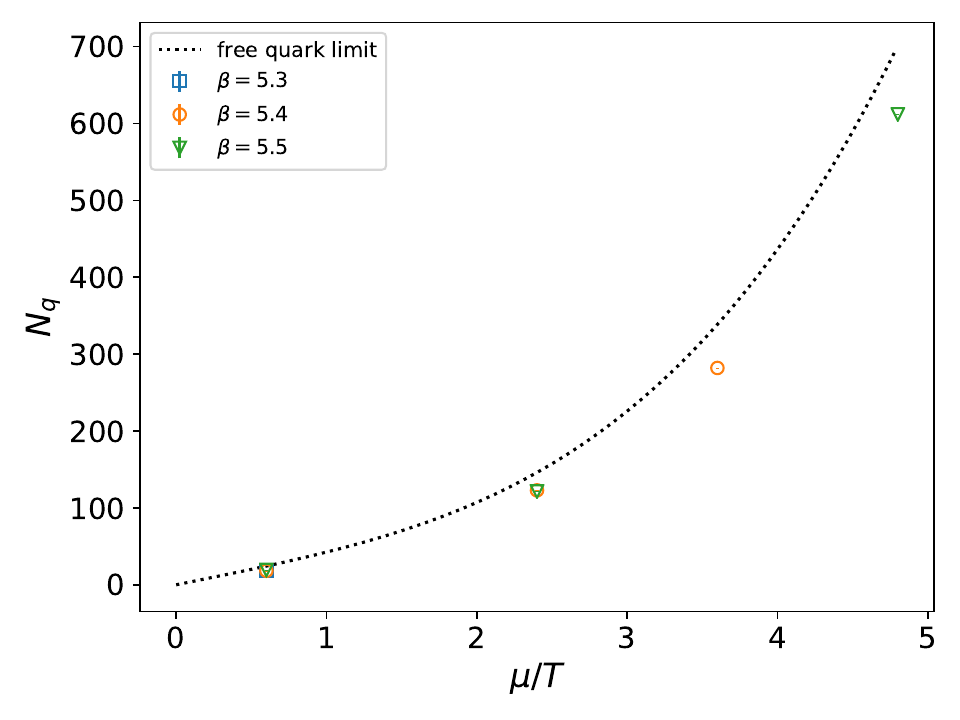}
	\includegraphics[width=8.0cm]{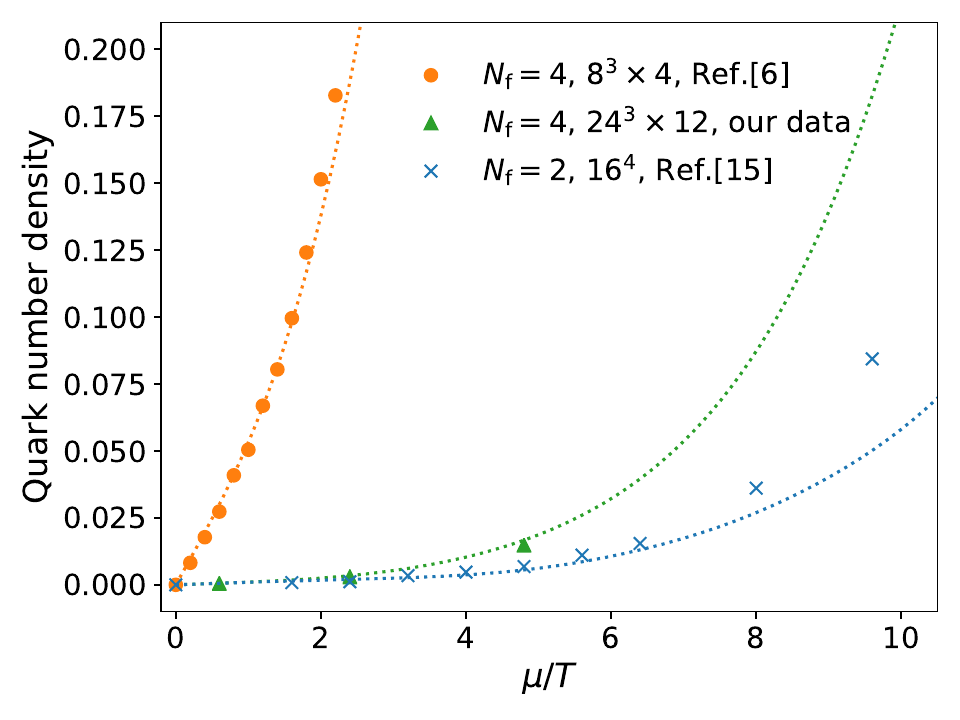}
	\caption{
	(Left) Quark number is plotted against $\mu / T$ at $5.2 \le \beta \le 5.5$.
		   The dotted line represents the result in the free-quark limit.
	           (Right) Quark number density obtained by the CL method
                   is plotted against $\mu / T$ for three different lattice setups.
                   The circles represent the results obtained in Ref.~\cite{Fodor:2015doa}
                   on a $8^3 \times 4$ lattice with four-flavor ($N_\text{f}=4$) staggered fermions
                   using $m=0.05$ at $\beta=5.4$. 
	           The triangles represent our results on a $24^3 \times 12$ lattice
                   with four-flavor ($N_\text{f}=4$) staggered fermions using $m=0.01$
                   at $\beta=5.5$, which are also plotted in the left panel.
	           The crosses represent the results obtained in Ref.~\cite{Kogut:2019qmi}
                   on a $16^4$ lattice with
                   two-flavor ($N_\text{f}=2$) staggered fermions using $m=0.025$
                   at $\beta=5.7$. The error bars are smaller than the symbols.
	The three dotted lines represent the free-quark results for each setup, which
        agree reasonably well with the corresponding data.}
	\label{fig:quark number at m=0.01}
\end{figure}

In Fig.~\ref{fig:quark number at m=0.01} (Left), we plot the quark number $N_{\rm q}$
against $\mu/T$ at $5.2 \le \beta \le 5.5$, where we find that $N_{\rm q}$ increases
monotonically with $\mu/T$, while it is almost independent of $\beta$.
This is understandable since $N_{\rm q}$ appears in the grand canonical partition
function $Z = {\rm Tr} e^{-(H+\mu N_{\rm q})/T}$ with the coefficient $\mu/T$.
In the same figure, we also plot the result for $N_{\rm q}$ in the free-quark limit,
which is given by
\begin{align}
N_{\rm q} 
&=
24\sum_{\bm{p}}
\left(
\frac{1}{e^{(E(\bm{p})-\mu)/T}}
-
\frac{1}{e^{(E(\bm{p})+\mu)/T}}
\right) \ ,
\label{free quark num}
\end{align}
using the single particle energy
\begin{align}
E(\bm{p}) &= \sinh^{-1}\sqrt{\sum_{i=1}^3 \sin^2(p_i) + m^2} \ ,
\quad  
p_i = \frac{2n_i \pi}{L_\mathrm{s}} \ ,
\quad 
-\frac{L_\mathrm{s}}{4} \leq n_i < \frac{L_\mathrm{s}}{4} \ .
\end{align}

\begin{figure}[H]
	\centering

	\includegraphics[width=8.0cm]{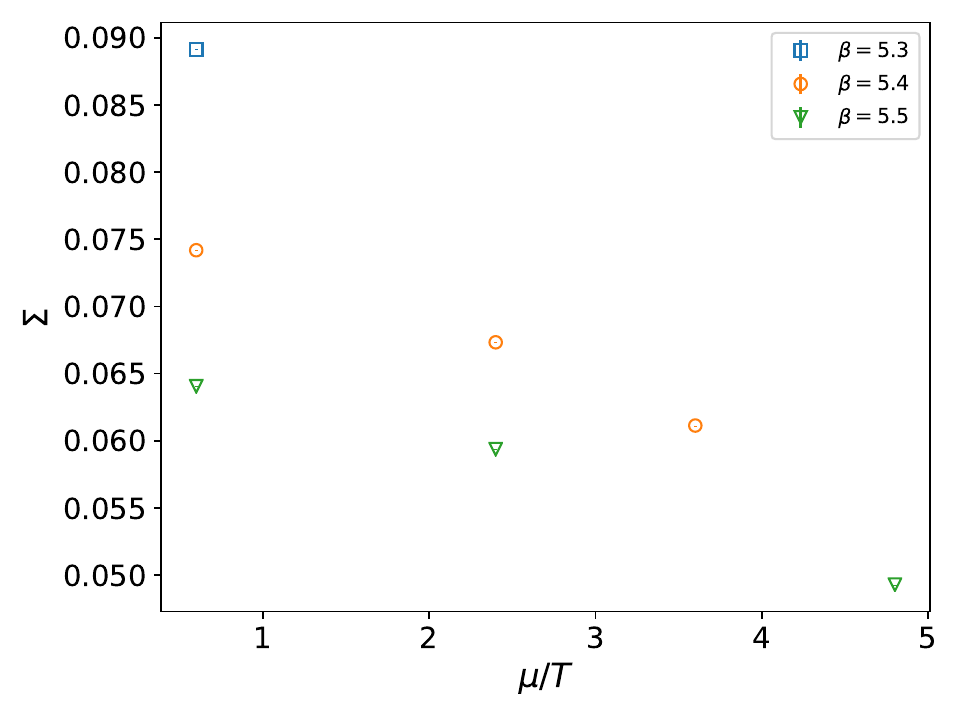}
	\includegraphics[width=8.0cm]{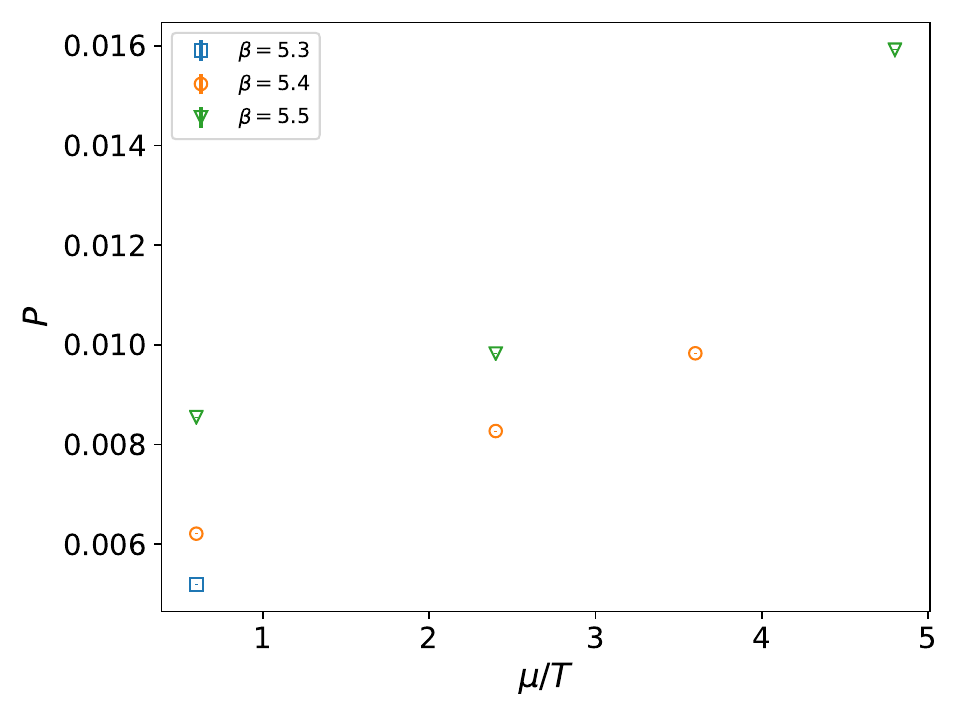}
	\caption{Chiral condensate $\Sigma$ (Left) and
		 the Polyakov loop $P$ (Right)          
		 are plotted against $\mu/T$ at various $\beta$ with $m = 0.01$.
                 The error bars are smaller than the size of the symbols.}
	\label{fig:order parameters at m=0.01}
\end{figure}

\clearpage

Let us recall that the CL results for $N_{\rm q}$ showed good agreement with the free-quark limit
in our previous study on $8^3 \times 16$ and $16^3 \times 32$ lattices
at $\beta = 5.7$~\cite{Ito:2020mys}.
This can be understood as a consequence of the fact that the spatial extent of
the lattice is $L = 0.34 {\rm ~fm}$ for $8^3 \times 16$ and $L = 0.68 {\rm ~fm}$
for $16^3 \times 32$, which are smaller than the typical QCD scale
$\Lambda_{\rm LQCD}^{-1}\sim  1{\rm ~fm}$, and hence the effective gauge coupling is actually small.
However, in our current simulation on a $24^3 \times 12$ lattice,
the spatial extent is $L = 1.3 - 2.7 {\rm ~fm}$ for $5.2 \le \beta \le 5.5$
and hence the effective gauge coupling is not small.
Despite this fact, the quark number
agrees reasonably well with the free-quark limit.
Similar behaviors were also observed in
previous results obtained by the CL method on an $8^3 \times 4$ lattice at $\beta=5.4$
with four-flavor staggered fermions using $m=0.05$ ~\cite{Fodor:2015doa}, and
those on a $16^4$ lattice at $\beta=5.7$ with two-flavor staggered fermions
using $m=0.025$~\cite{Kogut:2019qmi},
as shown in Fig.~\ref{fig:quark number at m=0.01} (Right).
This is in contrast to the situation with the equation of state, which agrees
with the free-quark limit given
by the Stefan-Boltzmann behavior
only at high temperature \cite{Engels:1996ag, Sexty:2019vqx}.

\section{Summary}\label{sec:summary}

We have investigated the validity region of the CL method
in finite density QCD with four-flavor staggered fermions
on a $24^3 \times 12$ lattice focusing on the deconfined phase near the
deconfining temperature.
We performed CL simulations at $\beta = 5.2 - 5.5$ and $\mu / T = 0.6 - 4.8$
with fixed bare quark mass $m = 0.01$.
The spatial extent is $L = (1.3 - 2.7 {\rm ~fm} )> \Lambda_{\rm LQCD}^{-1} \sim  1{\rm ~fm}$,
in contrast to our previous study at low-temperature high-density regime
performed with $L < \Lambda_{\rm LQCD}^{-1}$~\cite{Ito:2020mys}.
In a wide region of the deconfined phase, the CL method turned out to be valid
satisfying the justification criterion \cite{Nagata:2015uga}
based on the histogram of the drift terms,
which showed an exponential decay.
However, the CL method tends to fail as one approaches the critical temperature
due to the singular drift problem caused by the near-zero eigenvalues of the Dirac operator.
This tendency can be understood qualitatively by considering the generalized Banks-Casher relation, which relates the chiral condensate to the accumulation of the near-zero eigenvalues of the Dirac operator.

In fact the CL method is valid also in the low-temperature high-density regime
according to our previous study on $8^3 \times 16$ and $16^3 \times 32$ lattices
at $\beta = 5.7$~\cite{Ito:2020mys}.
By solving the Schwinger-Dyson equation
on a $8^3 \times 128$ lattice \cite{Yokota:2023osv},
the color superconducting phase was suggested to appear
even at large $\beta$
around discrete values of the chemical potential for which the Fermi surface
crosses the lattice quark momenta discretized due to finite size effects.
We are currently performing CL simulations to confirm this prediction \cite{Miura:2025aaa}.

Although the present study does not accumulate sufficient statistics to allow for precise discussions of physical observables, we have succeeded in narrowing down the parameter region where the complex Langevin method can be reliably applied. Increasing the statistics and investigating phenomena such as phase transitions at finite density remain important directions for future work.

\section*{Acknowledgments}
The authors are grateful to Kohtaroh Miura and Takeru Yokota for collaboration
in our ongoing projects related to the present work. This work is in part based on Bridge++ code (http://bridge.kek.jp/Lattice-code/).
This research was supported by MEXT as ``Priority Issue on Post-K computer''
(Elucidation of the Fundamental Laws and Evolution of the Universe) and
as ``Program for Promoting Researches on the Supercomputer Fugaku''
(Simulation for basic science: from fundamental laws of particles to creation of nuclei).
It is also supported by Joint Institute for Computational Fundamental Science (JICFuS).
Computations were carried out using computational resources of the K computer provided
by the RIKEN Center for Computational Science through the HPCI System Research
project (Project ID:hp180178, hp190159) and the Oakbridge-CX provided
by the Information Technology Center at the University of Tokyo through
the HPCI System Research project (Project ID:hp200079, hp210078, hp220094).
Y.A.\ was supported in part by JSPS KAKENHI Grant Number JP24K07036.
H.M., J.N.\ and Y.I.\ were supported in part by JSPS KAKENHI
Grant Numbers JP16H03988 and JP22H01224.
Y.N.\ was supported in part by JSPS KAKENHI
Grant Numbers JP18K03638, JP21K03553 and JP24K07052.
S.S.\ was supported by the MEXT-Supported Program for the Strategic Research Foundation
at Private Universities ``Topological Science'' (Grant No. S1511006).
S.T.\ was supported by the RIKEN Special Postdoctoral Researchers Program.

\appendix

\section{Review on the generalized Banks-Casher relation}
\label{sec:Banks-Casher}

In this appendix, we briefly review the generalized Banks-Casher relation,
which provides a qualitative understanding of our observation
in section \ref{sec:validity}
that the CL method tends to fail as we approach the critical temperature
of the deconfining phase transition.

In QCD at $\mu = 0$, it is well known that the chiral condensate
is given by
\begin{align}
\Sigma(m)
= \frac{1}{n} \expval{\Tr \frac{1}{D + m}}
= \frac{1}{n} \expval{\sum_{k=1}^n  \frac{1}{\im\lambda_k + m}} \ ,
\label{derive-Banks-Casher}
\end{align}
where $\expval{ \ \cdot \ }$ represents the expectation value with respect to
the partition function,
and $\{\im\lambda_k \}, \, (\lambda_k \in \mathbb{R}, \ k=1,\cdots , n)$ represent
the eigenvalues of the Dirac operator $D$.
Introducing the spectral density of the Dirac operator by
\begin{align}
\rho(\lambda) = \frac{1}{n} \expval{\sum_{k=1}^n  \delta(\lambda - \lambda_k)} \ ,
\end{align}
and rewriting the right-hand side of \eqref{derive-Banks-Casher} in terms of $\rho$,
one can prove the identity known as the Banks-Casher relation~\cite{Banks:1979yr},
\begin{align}
\lim_{m \to +0} \lim_{V \to \infty} \Sigma(m) = \pi \lim_{m\to +0} \lim_{V \to \infty} \rho(0) \ .
\label{eq:Banks-Casher}
\end{align}
Namely, the chiral condensate is dominated by the contribution from
low-lying eigenvalues.

This relation can be generalized to the case
with $\mu \neq 0$~\cite{Splittorff:2014zca,Nagata:2016alq}.
The Dirac operator is replaced by $(D + \mu \gamma_4)$, which is no longer anti-Hermitian.
Similarly to \eqref{derive-Banks-Casher}, the chiral condensate can be expressed as
\begin{align}
  \Sigma(m)
&= \frac{1}{n} \expval{\Tr \frac{1}{D + \mu \gamma_4 + m}}
  = \frac{1}{n} \expval{\sum_{k=1}^n  \frac{1}{x_k + \im y_k + m}}_\text{CL} \ ,
  \label{eq:Sigma-CL}
\end{align}
where $\expval{\ \cdot \ }_\text{CL}$ represents the Langevin time average in the CL simulation,
and $\{x_k + \im y_k\}, \, (x_k, y_k \in \mathbb{R})$
represent the eigenvalues of $(D + \mu \gamma_4)$.
The spectral density of $(D + \mu \gamma_4)$ can be defined as
\begin{align}
\rho_\text{CL}(x,y) &= \frac{1}{n} \expval{\sum_{k=1}^n \delta(x - x_k)\, \delta(y - y_k)}_\text{CL} \ .
\label{eq:CL spectral function}
\end{align}
Let us introduce the polar coordinates by $x = r\cos\theta$ and $y = r\sin\theta$, where $-\infty < r < \infty$ and $0 \le \theta < \pi$, and the corresponding distribution function
$\tilde{\rho}_\text{CL}(r,\theta) $.
The distribution functions $\rho_\text{CL}$ and $\tilde{\rho}_\text{CL}$ are positive semi-definite by definition. 
Rewriting the right-hand side of \eqref{eq:Sigma-CL} in terms of $\tilde{\rho}_\text{CL}$,
one can show the identity \cite{Nagata:2016alq}
\begin{align}
  \lim_{m \to +0}\lim_{V \to \infty}\Sigma(m) & =
  \pi  \lim_{r \to 0} \lim_{m \to +0} \lim_{V \to \infty}  r \hat{\rho}_\text{CL}(r) \ ,
\label{eq:generalized Banks-Casher}
  \\
\hat{\rho}_\text{CL}(r) & = \int_0^\pi \dd{\theta} \sin\theta \, \tilde{\rho}_\text{CL}(r, \theta) \ ,
\end{align}
which is a generalization of the Banks-Casher relation to finite density.
Explicit confirmation of \eqref{eq:generalized Banks-Casher} is given
in
Random Matrix Theory for finite density QCD \cite{Nagata:2016alq}.

Note that the distribution function \eqref{eq:CL spectral function} is
associated with the CL method.
In particular, they should not be confused with the eigenvalue distribution of the Dirac operator 
\begin{align}
  \rho(x,y) = \frac{1}{n}\expval{\sum_{k=1}^n \delta(x - x_k) \, \delta(y - y_k)}
\label{eq:complex-rho-DiracEV}
\end{align}
defined with respect to the original partition function,
which is complex-valued due to the complex weight.
The two definitions of the distribution are inequivalent even if the CL method is valid.
This is not a contradiction since the eigenvalue distribution of the Dirac operator is not a holomorphic quantity.
In other words, the distribution \eqref{eq:CL spectral function} defined by the CL method does not have a counterpart in
the original partition function.
In fact, it has been shown
in Ref.~\cite{Nagata:2016alq} 
that the distribution \eqref{eq:CL spectral function} can be changed by adopting a different gauge cooling criterion.
The quantity on the right-hand side of \eqref{eq:generalized Banks-Casher} cannot be changed, however, since it is related to the chiral condensate, which is a holomorphic quantity.

For the latter distribution, the relationship to the chiral condensate becomes more intricate \cite{Osborn:2005ss}.
All the eigenmodes contribute to the chiral condensate at finite density due to the violent phase oscillation of the fermion determinant.
In contrast, the generalized Banks-Casher relation \eqref{eq:generalized Banks-Casher} shows that only the low-lying eigenmodes of $(D + \mu \gamma_4)$ contribute to the chiral condensate
for configurations generated by the CL method.

Let us note here that the low-lying eigenmodes cause the singular drift problem since the fermionic part of the drift term \eqref{def-drift-fermi} involves the inverse of $M \sim (D + \mu \gamma_4 + m)$ as
\begin{align}
  v_{x,\nu}^\text{(f)}  &= - \Tr
 \left[ M^{-1}  \sum_{a=1}^8 \lambda_a
\left.\dv{\alpha} M(e^{\im \alpha \lambda_a}\mathcal{U}_{x,\nu}) \right|_{\alpha=0} \right] \ .
\label{drift and eigenmodes}
\end{align}
Thus the generalized Banks-Casher relation \eqref{eq:generalized Banks-Casher}
explains qualitatively our observation that the singular drift problem tends
to occur in the CL simulation as one approaches the chiral-symmetry-breaking confined phase.

\section{Thermalization and autocorrelation}
\label{app:thermalization}
To support the choice of thermalization time and the treatment of autocorrelation in our analysis, we provide supplementary results for the Polyakov loop obtained in simulations at $\beta = 5.3$, $5.4$, and $5.5$ with $\mu/T = 0.6$.

\begin{figure}[htbp]
    \centering
    \includegraphics[width=0.45\textwidth]{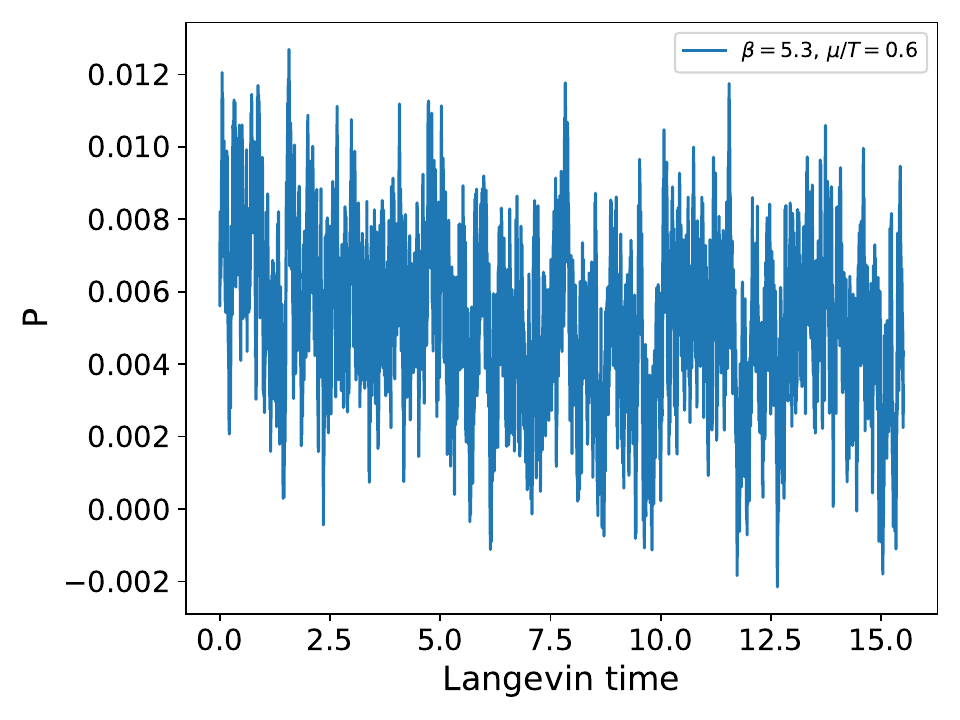}
    \includegraphics[width=0.45\textwidth]{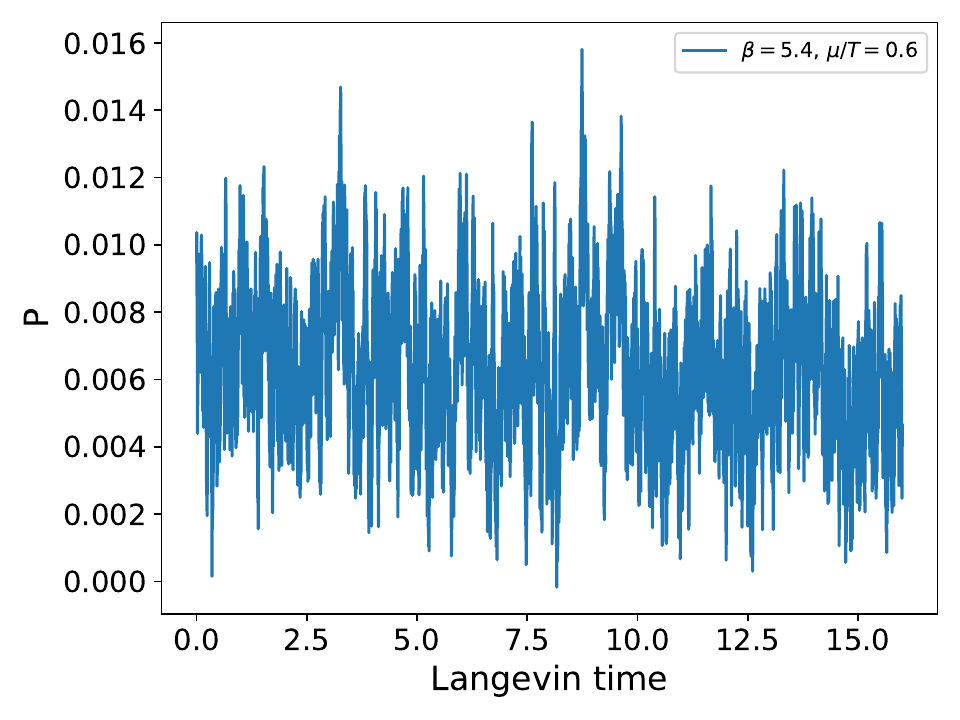}
    \includegraphics[width=0.45\textwidth]{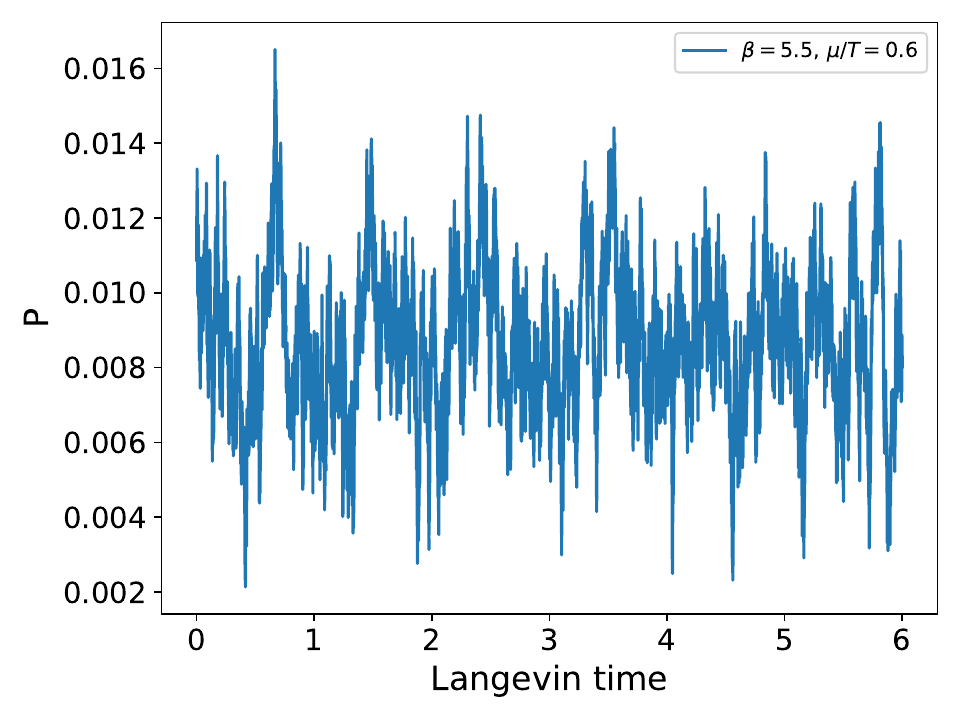}
    \caption{Langevin-time history of the Polyakov loop at $\mu/T = 0.6$ for $\beta = 5.3$ (Top-Left), $\beta = 5.4$ (Top-Right), and $\beta = 5.5$ (Bottom).}
    \label{fig:polyakov_history}
\end{figure}

\begin{figure}[htbp]
    \centering
    \includegraphics[width=0.45\textwidth]{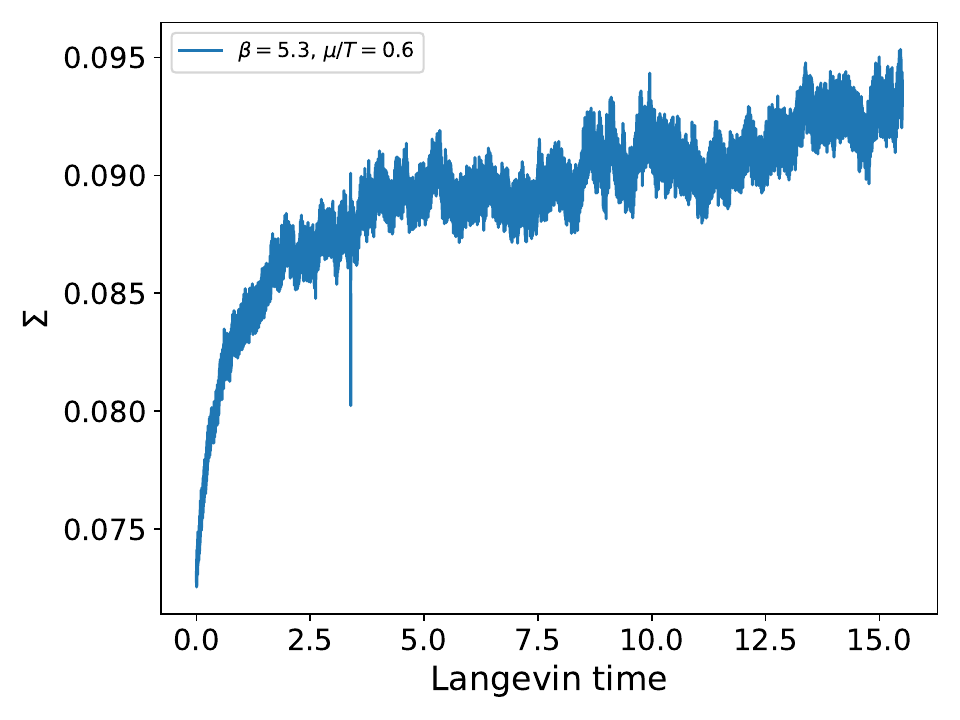}
    \includegraphics[width=0.45\textwidth]{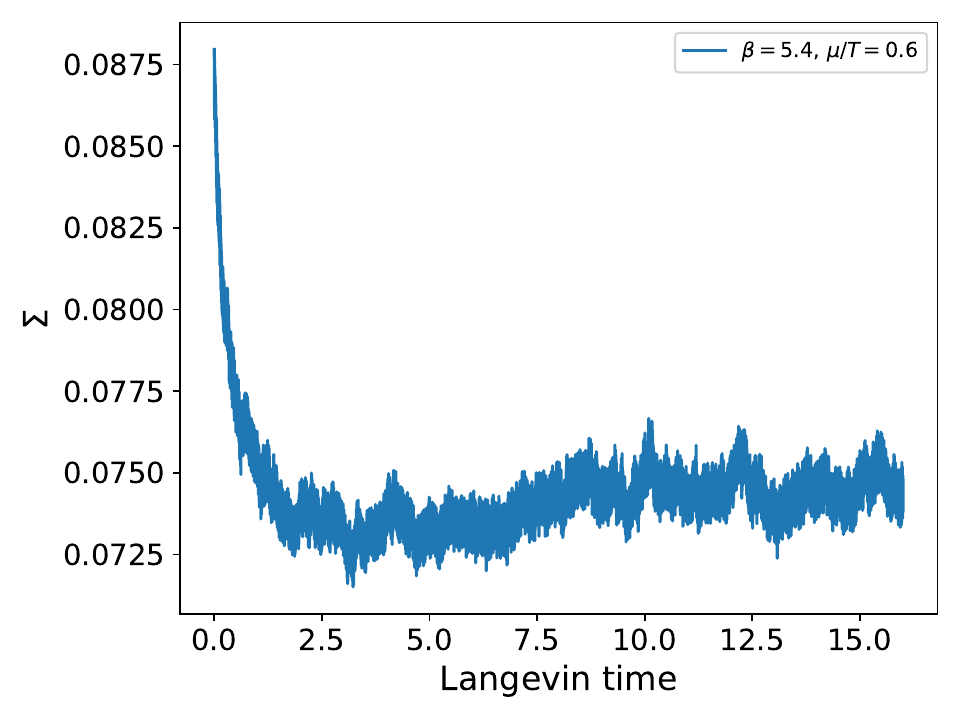}
    \includegraphics[width=0.45\textwidth]{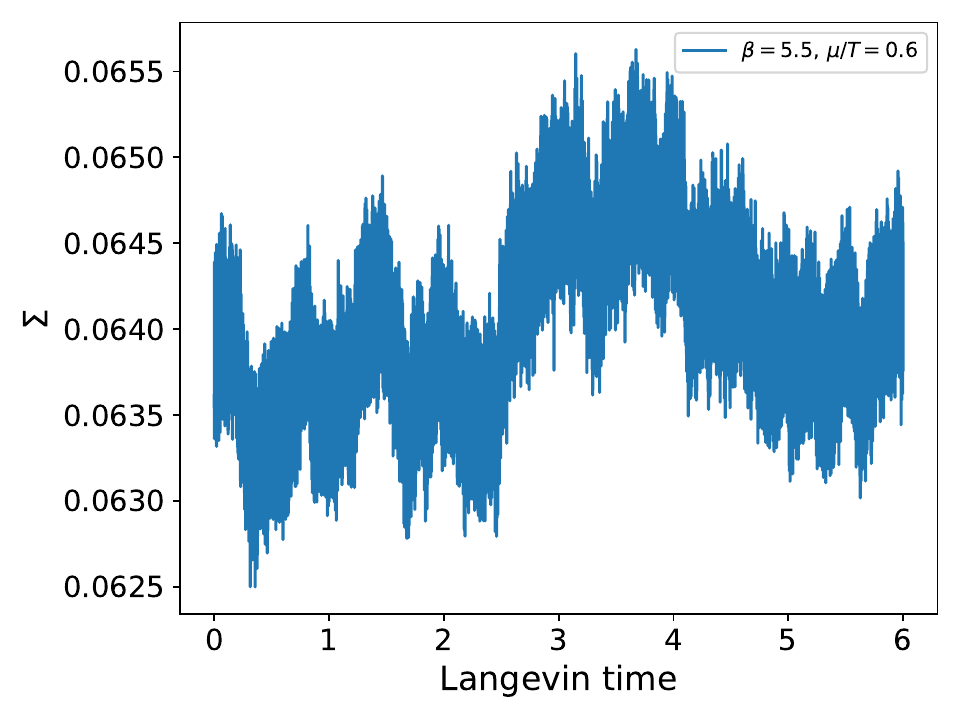}
    \caption{Langevin-time history of the chiral condensate at $\mu/T = 0.6$ for $\beta = 5.3$ (Top-Left), $\beta = 5.4$ (Top-Right), and $\beta = 5.5$ (Bottom).}
    \label{fig:chiral_history}
\end{figure}

\begin{figure}[htbp]
    \centering
    \includegraphics[width=0.45\textwidth]{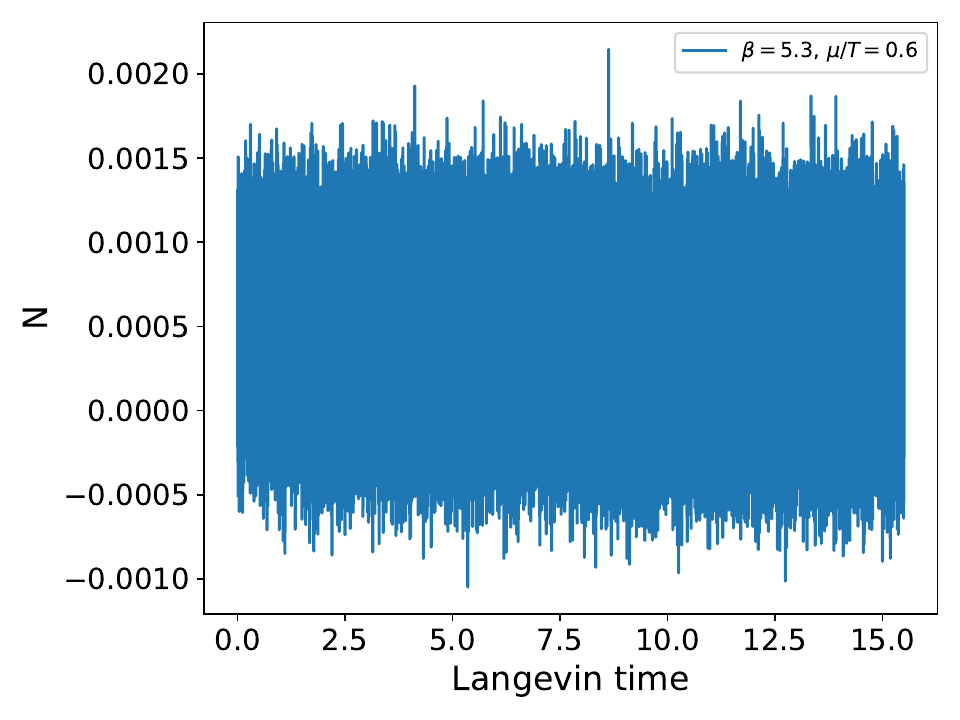}
    \includegraphics[width=0.45\textwidth]{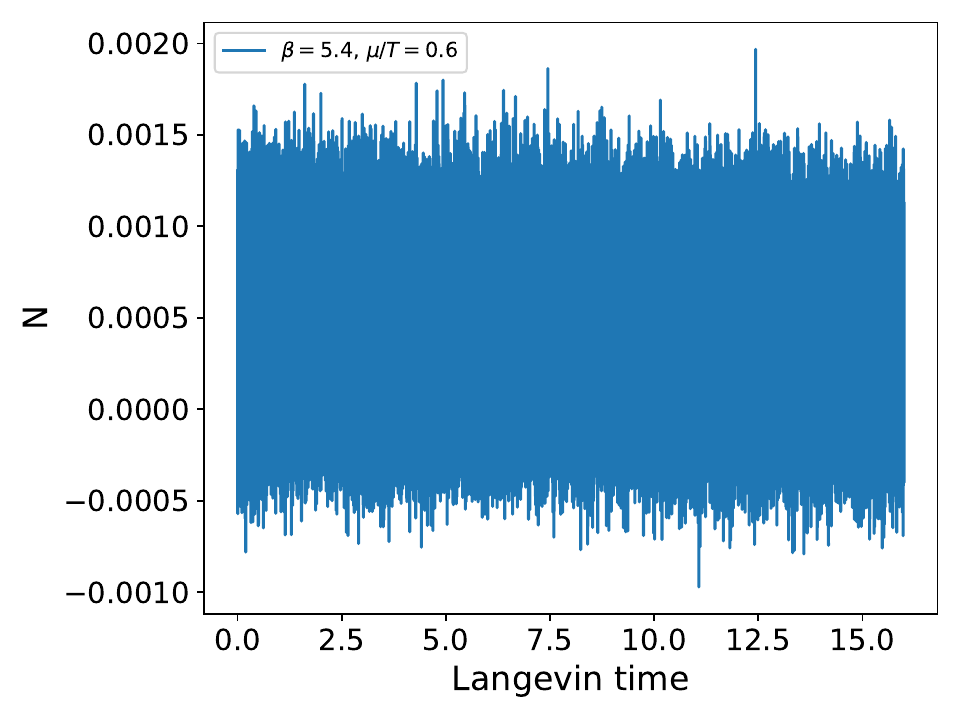}
    \includegraphics[width=0.45\textwidth]{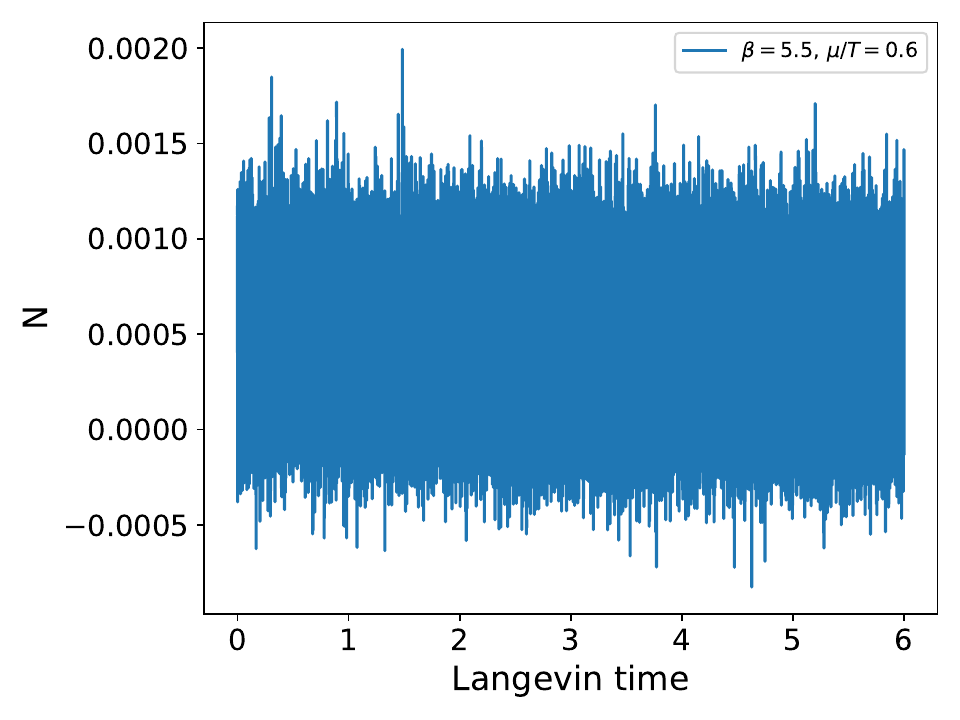}
    \caption{Langevin-time history of the baryon number density at $\mu/T = 0.6$ for $\beta = 5.3$ (Top-Left), $\beta = 5.4$ (Top-Right), and $\beta = 5.5$ (Bottom).}
    \label{fig:baryon_history}
\end{figure}

Figures~\ref{fig:polyakov_history},~\ref{fig:chiral_history} and \ref{fig:baryon_history} show the Langevin-time history of the Polyakov loop, the chiral condensate and the baryon number density $N = N_\text{q}/3V$ calculated at every Langevin step without thinning, providing a detailed account of the early-time dynamics.
For $\beta = 5.4$ and $5.5$, the Langevin-time histories do not exhibit clear transient behavior (except the chiral condensate), yet we conservatively discarded the initial region up to Langevin time $t = 4$ in order to minimize potential contamination from unthermalized configurations. In contrast, the ensemble at $\beta = 5.3$ shows a transient in the early Langevin evolution. To ensure sufficient thermalization in this case, we applied a stricter criterion and discarded data up to $t = 8$. We note that these differences in transient behavior may be attributed to differences in the initial configurations used in each simulation.

\begin{figure}[htbp]
    \centering
    \includegraphics[width=0.45\textwidth]{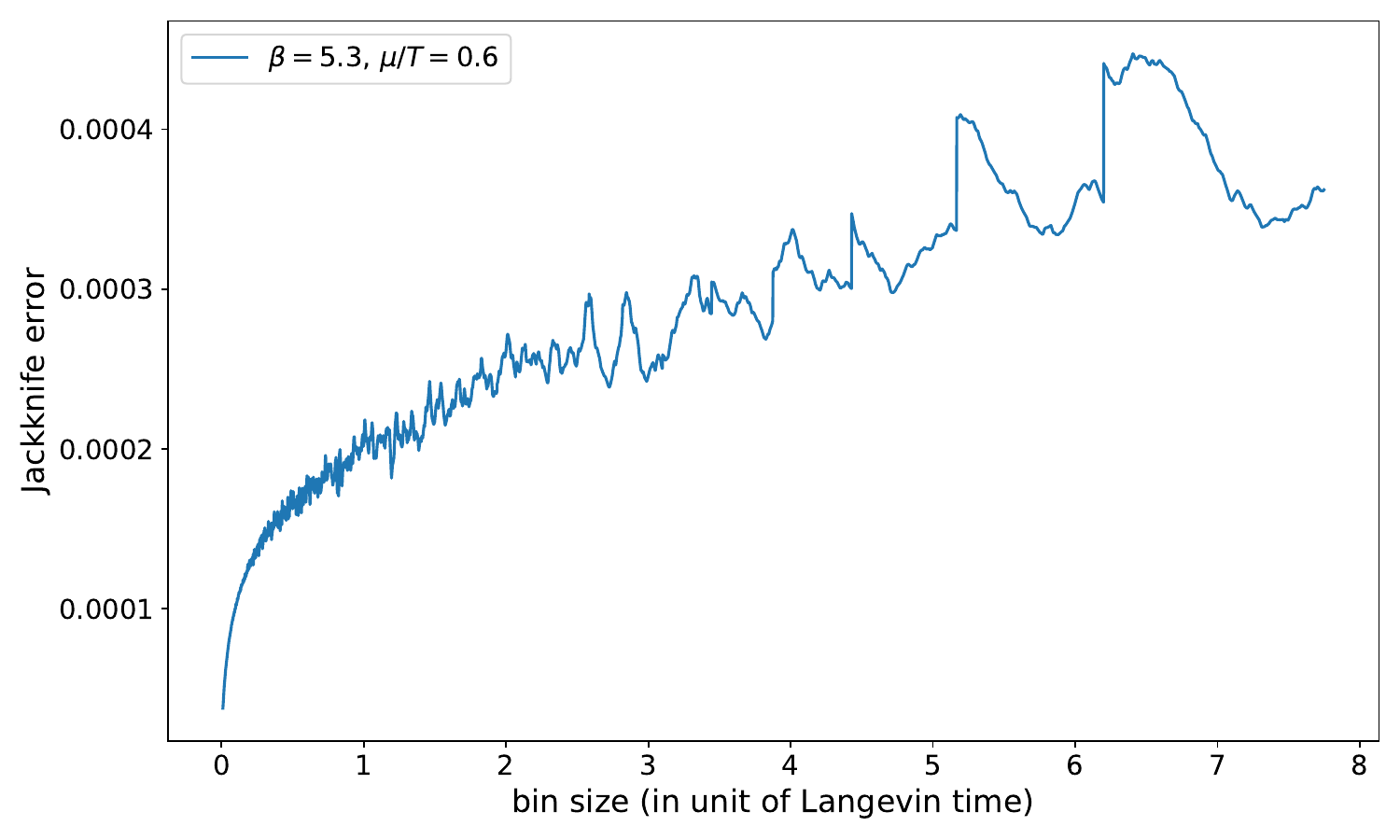}
    \includegraphics[width=0.45\textwidth]{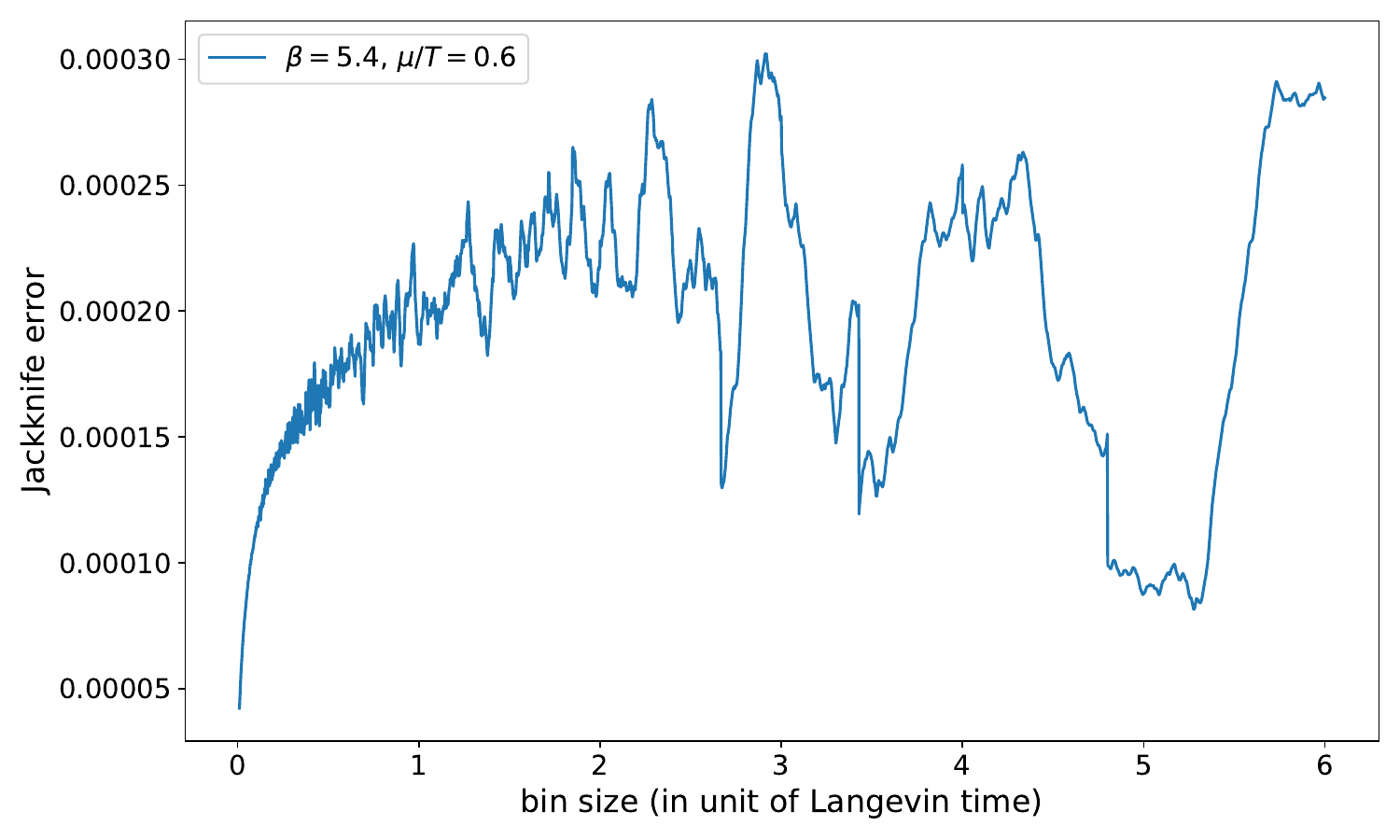}
    \includegraphics[width=0.45\textwidth]{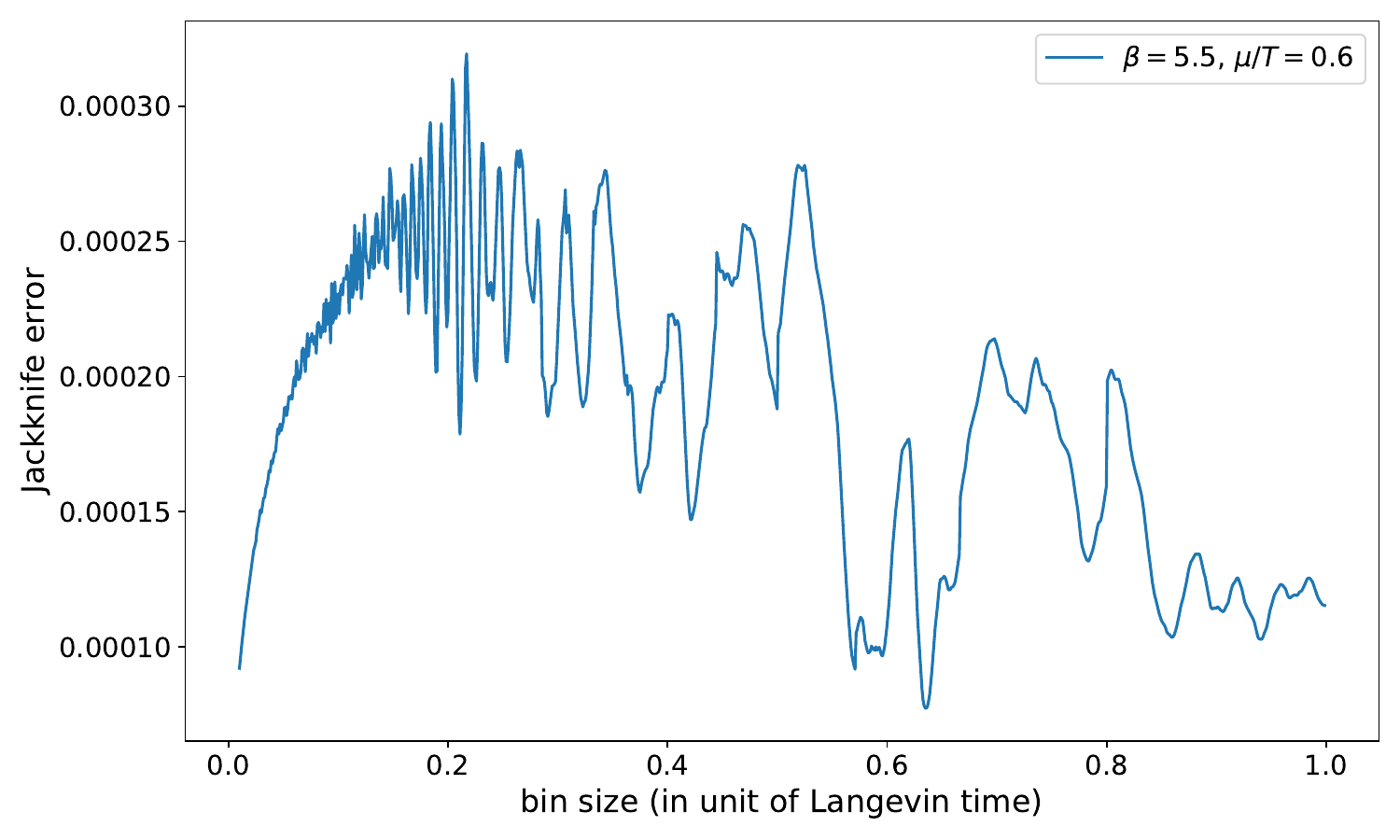}
    \caption{Jackknife error of the Polyakov loop as a function of bin size for the same ensembles as in Fig.~\ref{fig:polyakov_history}. The bin size is shown in units of Langevin time.}
    \label{fig:jackknife_binsize}
\end{figure}

\begin{figure}[htbp]
    \centering
    \includegraphics[width=0.45\textwidth]{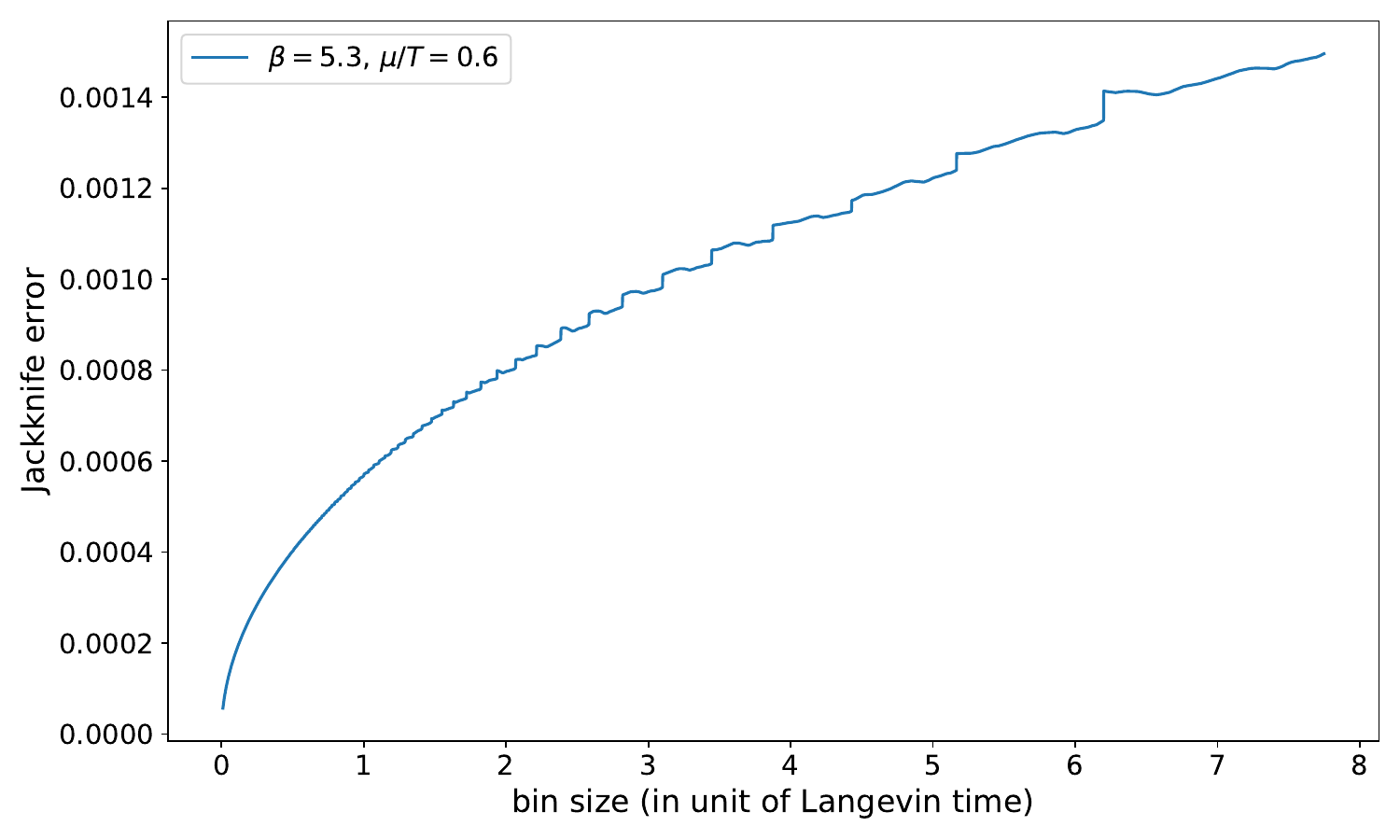}
    \includegraphics[width=0.45\textwidth]{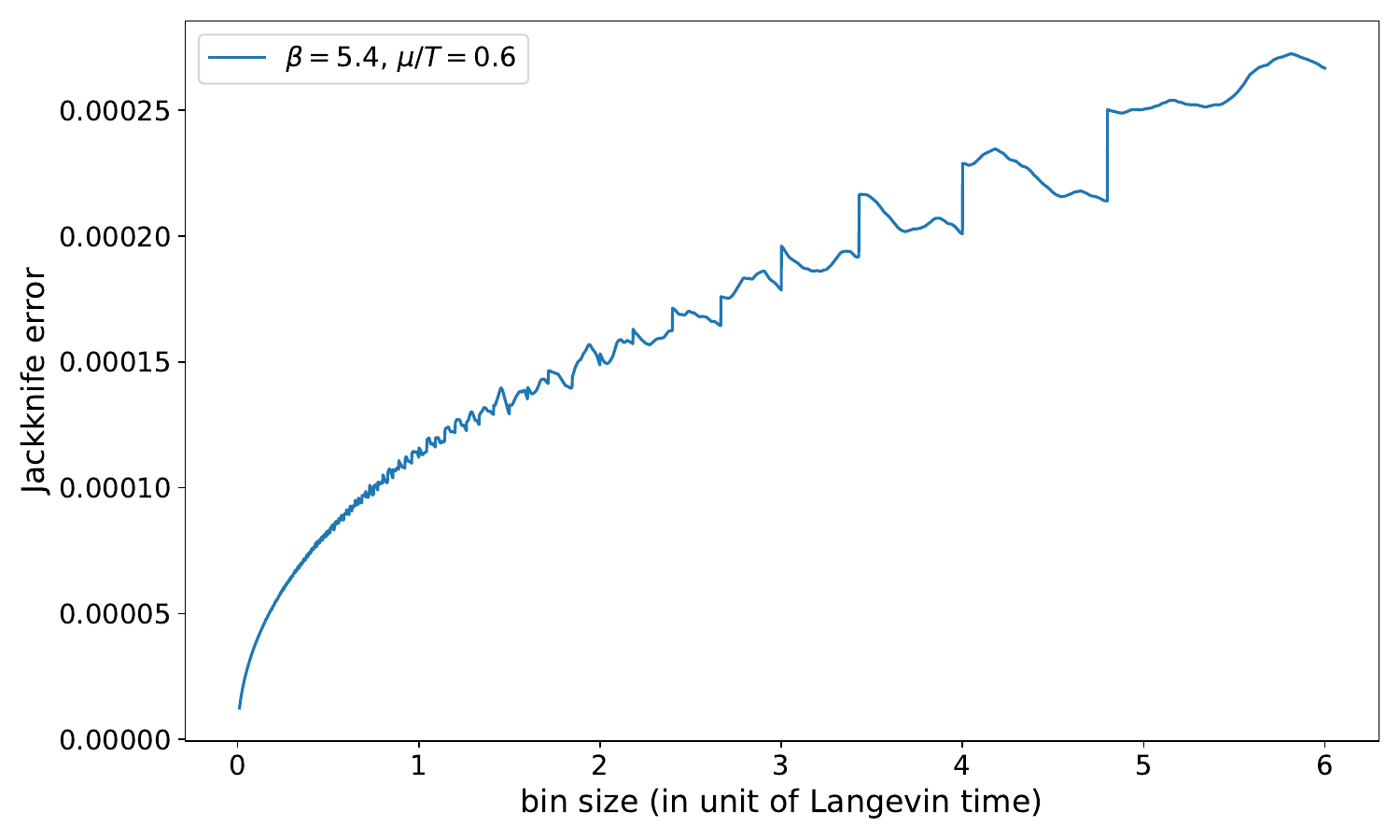}
    \includegraphics[width=0.45\textwidth]{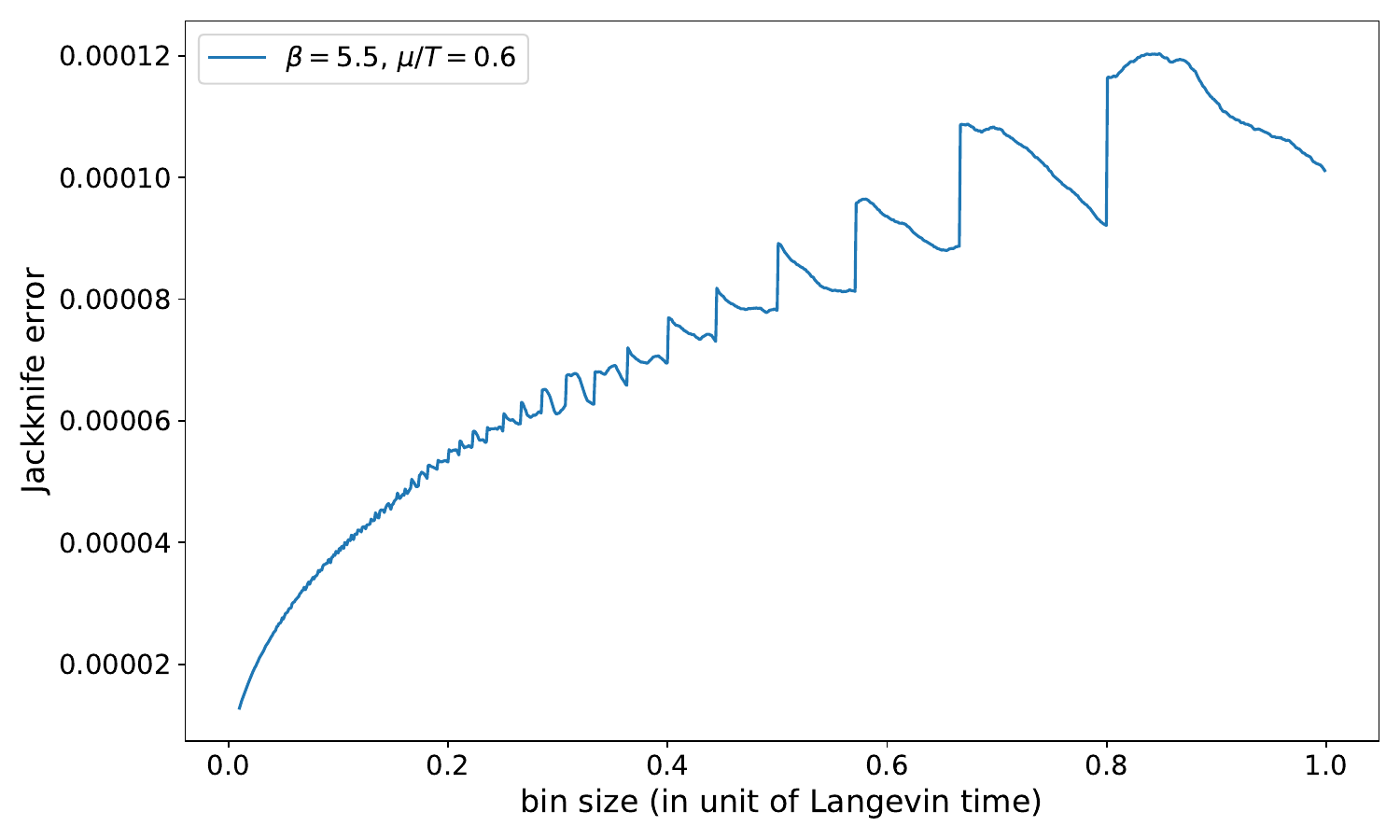}
    \caption{Jackknife error of the chiral condensate as a function of bin size for the same ensembles as in Fig.~\ref{fig:chiral_history}. The bin size is shown in units of Langevin time.}
    \label{fig:jackknife_binsize_chiral}
\end{figure}

\begin{figure}[htbp]
    \centering
    \includegraphics[width=0.45\textwidth]{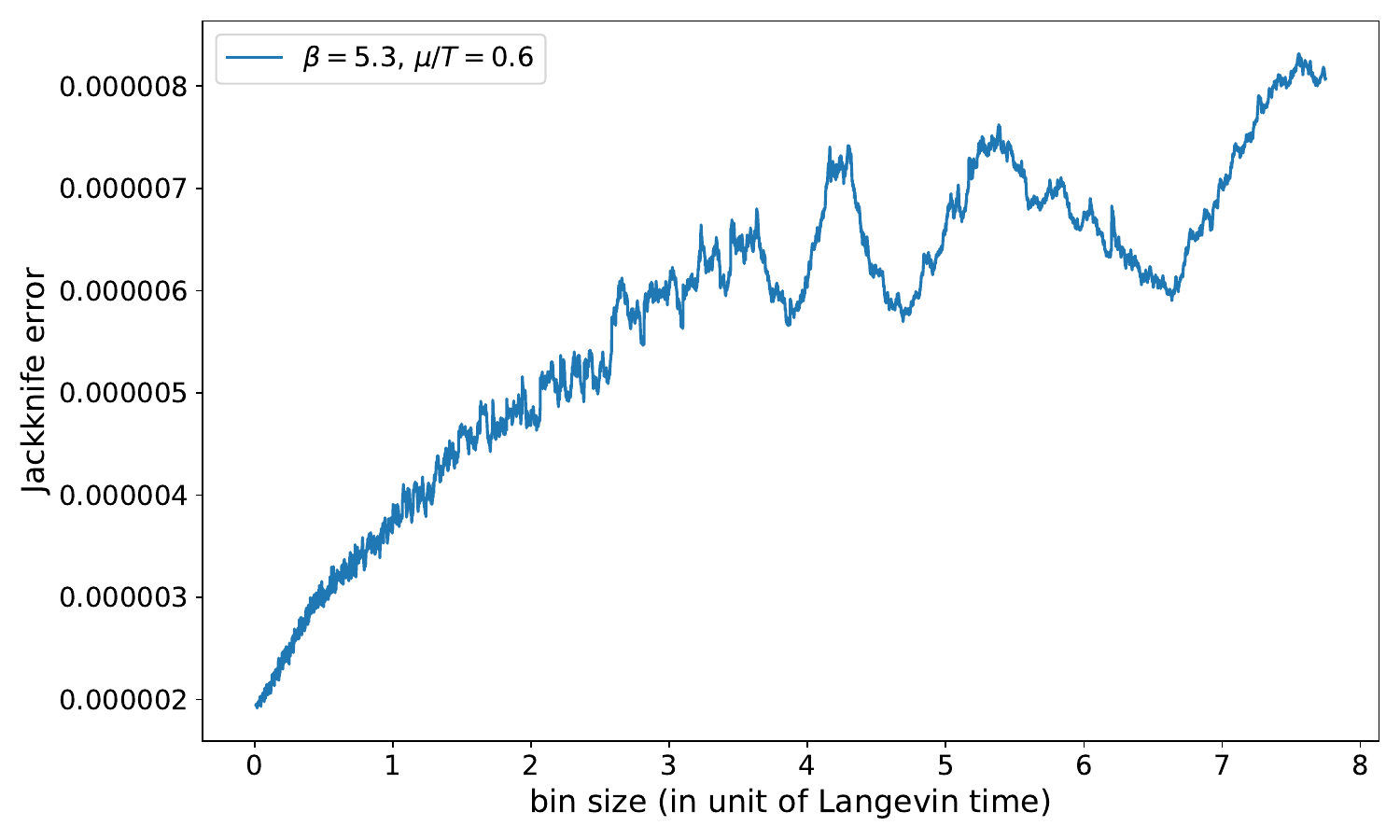}
    \includegraphics[width=0.45\textwidth]{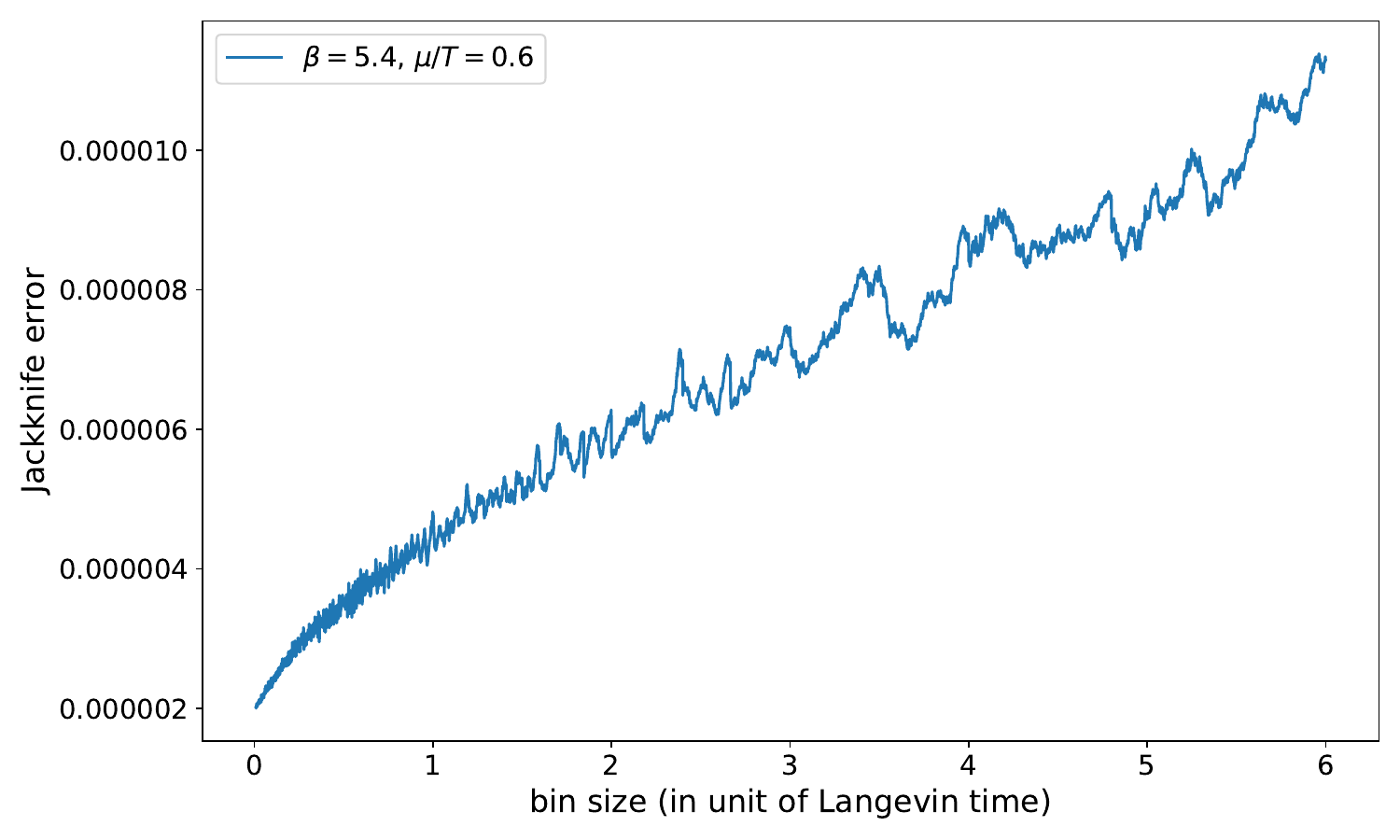}
    \includegraphics[width=0.45\textwidth]{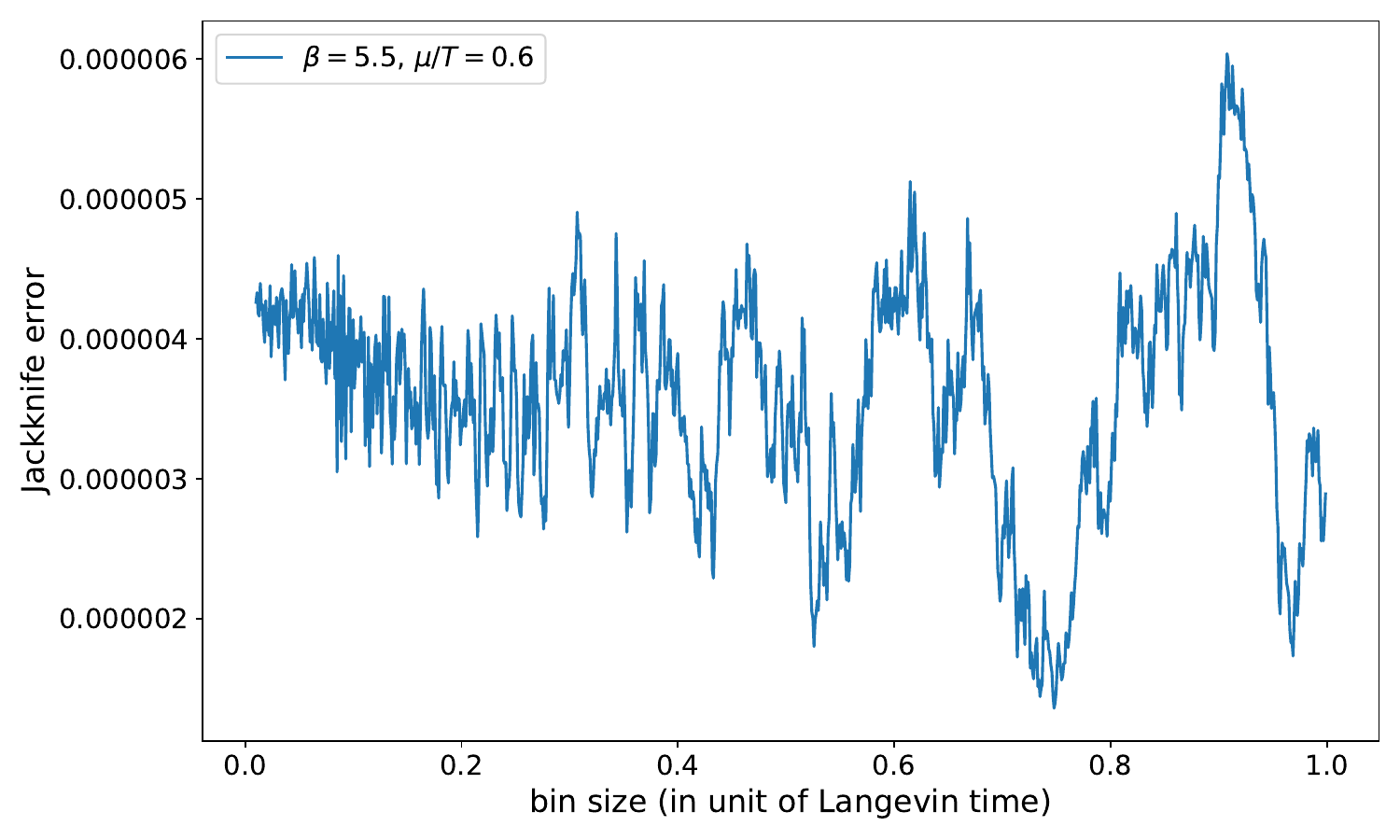}
    \caption{Jackknife error of the baryon number density as a function of bin size for the same ensembles as in Fig.~\ref{fig:baryon_history}. The bin size is shown in units of Langevin time.}
    \label{fig:jackknife_binsize_baryon}
\end{figure}

To evaluate autocorrelation effects on statistical error estimates, we investigated the dependence of the jackknife error on bin size. In this analysis, we constructed time series data by extracting every 10th Langevin step to reduce computational cost while preserving temporal structure. 
The results of the jackknife analysis indicate that the autocorrelation time is relatively long for the ensembles at $\beta = 5.3$ and $5.4$, as shown in Figures~\ref{fig:jackknife_binsize},~\ref{fig:jackknife_binsize_chiral} and~\ref{fig:jackknife_binsize_baryon}. In particular, the jackknife error continues to increase with the bin size over a wide range, suggesting that many of the configurations are not statistically independent. This implies that the number of effectively independent samples is significantly reduced in these ensembles, and care must be taken when interpreting the statistical uncertainties. 



\clearpage

\bibliographystyle{JHEP}
\bibliography{ref.bib}

\providecommand{\href}[2]{#2}\begingroup\raggedright\begin{thebibliography}{10}

\bibitem{Ito:2020mys}
Y.~Ito, H.~Matsufuru, Y.~Namekawa, J.~Nishimura, S.~Shimasaki, A.~Tsuchiya, and
  S.~Tsutsui, {\it {Complex Langevin calculations in QCD at finite density}},
  {\em JHEP} {\bf 10} (2020) 144, [\href{http://arxiv.org/abs/2007.08778}{{\tt
  arXiv:2007.08778}}].

\bibitem{Klauder:1983sp}
J.~R. Klauder, {\it {Coherent state Langevin equations for canonical quantum
  systems with applications to the quantized Hall effect}},  {\em Phys. Rev.}
  {\bf A29} (1984) 2036--2047.

\bibitem{Parisi:1984cs}
G.~Parisi, {\it {On complex probabilities}},  {\em Phys. Lett.} {\bf 131B}
  (1983) 393--395.

\bibitem{Sexty:2013ica}
D.~Sexty, {\it {Simulating full QCD at nonzero density using the complex
  Langevin equation}},  {\em Phys. Lett.} {\bf B729} (2014) 108--111,
  [\href{http://arxiv.org/abs/1307.7748}{{\tt arXiv:1307.7748}}].

\bibitem{Aarts:2014bwa}
G.~Aarts, E.~Seiler, D.~Sexty, and I.-O. Stamatescu, {\it {Simulating QCD at
  nonzero baryon density to all orders in the hopping parameter expansion}},
  {\em Phys. Rev.} {\bf D90} (2014), no.~11 114505,
  [\href{http://arxiv.org/abs/1408.3770}{{\tt arXiv:1408.3770}}].

\bibitem{Fodor:2015doa}
Z.~Fodor, S.~Katz, D.~Sexty, and C.~Torok, {\it {Complex Langevin dynamics for
  dynamical QCD at nonzero chemical potential: A comparison with multiparameter
  reweighting}},  {\em Phys. Rev. D} {\bf 92} (2015), no.~9 094516,
  [\href{http://arxiv.org/abs/1508.05260}{{\tt arXiv:1508.05260}}].

\bibitem{Sinclair:2015kva}
D.~K. Sinclair and J.~Kogut, {\it {Exploring complex-Langevin methods for
  finite-density QCD}},  {\em PoS} {\bf LATTICE2015} (2016) 153,
  [\href{http://arxiv.org/abs/1510.06367}{{\tt arXiv:1510.06367}}].

\bibitem{Sinclair:2016nbg}
D.~Sinclair and J.~Kogut, {\it {Complex Langevin for lattice QCD at $T=0$ and
  $\mu \ge 0$}},  {\em PoS} {\bf LATTICE2016} (2016) 026,
  [\href{http://arxiv.org/abs/1611.02312}{{\tt arXiv:1611.02312}}].

\bibitem{Sinclair:2017zhn}
D.~Sinclair and J.~Kogut, {\it {Complex Langevin simulations of QCD at finite
  density -- Progress report}},  {\em EPJ Web Conf.} {\bf 175} (2018) 07031,
  [\href{http://arxiv.org/abs/1710.08465}{{\tt arXiv:1710.08465}}].

\bibitem{Sinclair:2018rbk}
D.~Sinclair and J.~Kogut, {\it {Complex Langevin for lattice QCD}},  {\em PoS}
  {\bf LATTICE2018} (2018) 143, [\href{http://arxiv.org/abs/1810.11880}{{\tt
  arXiv:1810.11880}}].

\bibitem{Nagata:2018mkb}
K.~Nagata, J.~Nishimura, and S.~Shimasaki, {\it {Complex Langevin calculations
  in finite density QCD at large $\mu/T$ with the deformation technique}},
  {\em Phys. Rev.} {\bf D98} (2018), no.~11 114513,
  [\href{http://arxiv.org/abs/1805.03964}{{\tt arXiv:1805.03964}}].

\bibitem{Ito:2018jpo}
Y.~Ito, H.~Matsufuru, J.~Nishimura, S.~Shimasaki, A.~Tsuchiya, and S.~Tsutsui,
  {\it {Exploring the phase diagram of finite density QCD at low temperature by
  the complex Langevin method}},  {\em PoS} {\bf LATTICE2018} (2018) 146,
  [\href{http://arxiv.org/abs/1811.12688}{{\tt arXiv:1811.12688}}].

\bibitem{Tsutsui:2018jva}
S.~Tsutsui, Y.~Ito, H.~Matsufuru, J.~Nishimura, S.~Shimasaki, and A.~Tsuchiya,
  {\it {Can the complex Langevin method see the deconfinement phase transition
  in QCD at finite density?}},  {\em PoS} {\bf LATTICE2018} (2018) 144,
  [\href{http://arxiv.org/abs/1811.07647}{{\tt arXiv:1811.07647}}].

\bibitem{Tsutsui:2019gwn}
S.~Tsutsui, Y.~Ito, H.~Matsufuru, J.~Nishimura, S.~Shimasaki, and A.~Tsuchiya,
  {\it {Exploring the finite density QCD based on the complex Langevin
  method}},  {\em JPS Conf. Proc.} {\bf 26} (2019) 024012.

\bibitem{Kogut:2019qmi}
J.~B. Kogut and D.~K. Sinclair, {\it {Applying complex Langevin simulations to
  lattice QCD at finite density}},  {\em Phys. Rev.} {\bf D100} (2019), no.~5
  054512, [\href{http://arxiv.org/abs/1903.02622}{{\tt arXiv:1903.02622}}].

\bibitem{Sexty:2019vqx}
D.~Sexty, {\it {Calculating the equation of state of dense quark-gluon plasma
  using the complex Langevin equation}},  {\em Phys. Rev. D} {\bf 100} (2019),
  no.~7 074503, [\href{http://arxiv.org/abs/1907.08712}{{\tt
  arXiv:1907.08712}}].

\bibitem{Sinclair:2019ysx}
D.~K. Sinclair and J.~B. Kogut, {\it {Applying complex Langevin to lattice QCD
  at finite $\mu$}},  in {\em {37th International Symposium on Lattice Field
  Theory (Lattice 2019) Wuhan, Hubei, China, June 16-22, 2019}},
  vol.~LATTICE2019, p.~245, 2019.
\newblock \href{http://arxiv.org/abs/1910.11412}{{\tt arXiv:1910.11412}}.

\bibitem{Tsutsui:2019suq}
S.~Tsutsui, Y.~Ito, H.~Matsufuru, J.~Nishimura, S.~Shimasaki, and A.~Tsuchiya,
  {\it {Exploring the QCD phase diagram at finite density by the complex
  Langevin method on a $16^3\times 32$ lattice}},  in {\em {37th International
  Symposium on Lattice Field Theory (Lattice 2019) Wuhan, Hubei, China, June
  16-22, 2019}}, 2019.
\newblock \href{http://arxiv.org/abs/1912.00361}{{\tt arXiv:1912.00361}}.

\bibitem{Scherzer:2020kiu}
M.~Scherzer, D.~Sexty, and I.~O. Stamatescu, {\it {Deconfinement transition
  line with the complex Langevin equation up to $\mu /T \sim 5$}},  {\em Phys.
  Rev. D} {\bf 102} (2020), no.~1 014515,
  [\href{http://arxiv.org/abs/2004.05372}{{\tt arXiv:2004.05372}}].

\bibitem{Berger:2019odf}
C.~E. Berger, L.~Rammelm\"uller, A.~C. Loheac, F.~Ehmann, J.~Braun, and J.~E.
  Drut, {\it {Complex Langevin and other approaches to the sign problem in
  quantum many-body physics}},  {\em Phys. Rept.} {\bf 892} (2021) 1--54,
  [\href{http://arxiv.org/abs/1907.10183}{{\tt arXiv:1907.10183}}].

\bibitem{Attanasio:2020spv}
F.~Attanasio, B.~J\"ager, and F.~P.~G. Ziegler, {\it {Complex Langevin
  simulations and the QCD phase diagram: Recent developments}},  {\em Eur.
  Phys. J. A} {\bf 56} (2020), no.~10 251,
  [\href{http://arxiv.org/abs/2006.00476}{{\tt arXiv:2006.00476}}].

\bibitem{Nagata:2021ugx}
K.~Nagata, {\it {Finite-density lattice QCD and sign problem: Current status
  and open problems}},  {\em Prog. Part. Nucl. Phys.} {\bf 127} (2022) 103991,
  [\href{http://arxiv.org/abs/2108.12423}{{\tt arXiv:2108.12423}}].

\bibitem{Parisi:1980ys}
G.~Parisi and Y.-s. Wu, {\it {Perturbation theory without gauge fixing}},  {\em
  Sci. Sin.} {\bf 24} (1981) 483.

\bibitem{Aarts:2009uq}
G.~Aarts, E.~Seiler, and I.-O. Stamatescu, {\it {The complex Langevin method:
  When can it be trusted?}},  {\em Phys. Rev.} {\bf D81} (2010) 054508,
  [\href{http://arxiv.org/abs/0912.3360}{{\tt arXiv:0912.3360}}].

\bibitem{Aarts:2011ax}
G.~Aarts, F.~A. James, E.~Seiler, and I.-O. Stamatescu, {\it {Complex Langevin:
  Etiology and diagnostics of its main problem}},  {\em Eur. Phys. J.} {\bf
  C71} (2011) 1756, [\href{http://arxiv.org/abs/1101.3270}{{\tt
  arXiv:1101.3270}}].

\bibitem{Nishimura:2015pba}
J.~Nishimura and S.~Shimasaki, {\it {New insights into the problem with a
  singular drift term in the complex Langevin method}},  {\em Phys. Rev.} {\bf
  D92} (2015), no.~1 011501, [\href{http://arxiv.org/abs/1504.08359}{{\tt
  arXiv:1504.08359}}].

\bibitem{Nagata:2015uga}
K.~Nagata, J.~Nishimura, and S.~Shimasaki, {\it {Justification of the complex
  Langevin method with the gauge cooling procedure}},  {\em PTEP} {\bf 2016}
  (2016), no.~1 013B01, [\href{http://arxiv.org/abs/1508.02377}{{\tt
  arXiv:1508.02377}}].

\bibitem{Aarts:2017vrv}
G.~Aarts, E.~Seiler, D.~Sexty, and I.-O. Stamatescu, {\it {Complex Langevin
  dynamics and zeroes of the fermion determinant}},  {\em JHEP} {\bf 05} (2017)
  044, [\href{http://arxiv.org/abs/1701.02322}{{\tt arXiv:1701.02322}}].
  [Erratum: JHEP01,128(2018)].

\bibitem{Scherzer:2018hid}
M.~Scherzer, E.~Seiler, D.~Sexty, and I.-O. Stamatescu, {\it {Complex Langevin
  and boundary terms}},  {\em Phys. Rev.} {\bf D99} (2019), no.~1 014512,
  [\href{http://arxiv.org/abs/1808.05187}{{\tt arXiv:1808.05187}}].

\bibitem{Scherzer:2019lrh}
M.~Scherzer, E.~Seiler, D.~Sexty, and I.~O. Stamatescu, {\it {Controlling
  complex Langevin simulations of lattice models by boundary term analysis}},
  {\em Phys. Rev.} {\bf D101} (2020), no.~1 014501,
  [\href{http://arxiv.org/abs/1910.09427}{{\tt arXiv:1910.09427}}].

\bibitem{Nagata:2016vkn}
K.~Nagata, J.~Nishimura, and S.~Shimasaki, {\it {Argument for justification of
  the complex Langevin method and the condition for correct convergence}},
  {\em Phys. Rev.} {\bf D94} (2016), no.~11 114515,
  [\href{http://arxiv.org/abs/1606.07627}{{\tt arXiv:1606.07627}}].

\bibitem{Nagata:2018net}
K.~Nagata, J.~Nishimura, and S.~Shimasaki, {\it {Testing the criterion for
  correct convergence in the complex Langevin method}},  {\em JHEP} {\bf 05}
  (2018) 004, [\href{http://arxiv.org/abs/1802.01876}{{\tt arXiv:1802.01876}}].

\bibitem{Salcedo:2016kyy}
L.~L. Salcedo, {\it {Does the complex Langevin method give unbiased results?}},
   {\em Phys. Rev.} {\bf D94} (2016), no.~11 114505,
  [\href{http://arxiv.org/abs/1611.06390}{{\tt arXiv:1611.06390}}].

\bibitem{Cai:2020tgd}
Z.~Cai, X.~Dong, and Y.~Kuang, {\it {On the validity of complex Langevin method
  for path integral computations}},  {\em SIAM J. Sci. Comput.} {\bf 43}
  (2021), no.~1 A685--A719, [\href{http://arxiv.org/abs/2007.10198}{{\tt
  arXiv:2007.10198}}].

\bibitem{Mollgaard:2013qra}
A.~Mollgaard and K.~Splittorff, {\it {Complex Langevin dynamics for chiral
  Random Matrix Theory}},  {\em Phys. Rev. D} {\bf 88} (2013), no.~11 116007,
  [\href{http://arxiv.org/abs/1309.4335}{{\tt arXiv:1309.4335}}].

\bibitem{Fukugita:1990vu}
M.~Fukugita, H.~Mino, M.~Okawa, and A.~Ukawa, {\it {Finite size test for the
  finite temperature chiral phase transition in lattice QCD}},  {\em Phys.\
  Rev.\ Lett.} {\bf 65} (1990) 816--819.

\bibitem{Engels:1996ag}
J.~Engels, R.~Joswig, F.~Karsch, E.~Laermann, M.~Lutgemeier, and B.~Petersson,
  {\it {Thermodynamics of four flavor QCD with improved staggered fermions}},
  {\em Phys.\ Lett.\ B} {\bf 396} (1997) 210--216,
  [\href{http://arxiv.org/abs/hep-lat/9612018}{{\tt hep-lat/9612018}}].

\bibitem{Fodor:2001au}
Z.~Fodor and S.~D. Katz, {\it {A new method to study lattice QCD at finite
  temperature and chemical potential}},  {\em Phys. Lett.} {\bf B534} (2002)
  87--92, [\href{http://arxiv.org/abs/hep-lat/0104001}{{\tt hep-lat/0104001}}].

\bibitem{DElia:2002tig}
M.~D'Elia and M.-P. Lombardo, {\it {Finite density QCD via imaginary chemical
  potential}},  {\em Phys. Rev.} {\bf D67} (2003) 014505,
  [\href{http://arxiv.org/abs/hep-lat/0209146}{{\tt hep-lat/0209146}}].

\bibitem{DElia:2004ani}
M.~D'Elia and M.~P. Lombardo, {\it {QCD thermodynamics from an imaginary
  $\mu(B)$: Results on the four flavor lattice model}},  {\em Phys.\ Rev.\ D}
  {\bf 70} (2004) 074509, [\href{http://arxiv.org/abs/hep-lat/0406012}{{\tt
  hep-lat/0406012}}].

\bibitem{Azcoiti:2004ri}
V.~Azcoiti, G.~Di~Carlo, A.~Galante, and V.~Laliena, {\it {Finite density QCD:
  A new approach}},  {\em JHEP} {\bf 12} (2004) 010,
  [\href{http://arxiv.org/abs/hep-lat/0409157}{{\tt hep-lat/0409157}}].

\bibitem{Azcoiti:2005tv}
V.~Azcoiti, G.~Di~Carlo, A.~Galante, and V.~Laliena, {\it {Phase diagram of QCD
  with four quark flavors at finite temperature and baryon density}},  {\em
  Nucl. Phys.} {\bf B723} (2005) 77--90,
  [\href{http://arxiv.org/abs/hep-lat/0503010}{{\tt hep-lat/0503010}}].

\bibitem{deForcrand:2006ec}
P.~de~Forcrand and S.~Kratochvila, {\it {Finite density QCD with a canonical
  approach}},  {\em Nucl. Phys. Proc. Suppl.} {\bf 153} (2006) 62--67,
  [\href{http://arxiv.org/abs/hep-lat/0602024}{{\tt hep-lat/0602024}}].

\bibitem{Fodor:2007vv}
Z.~Fodor, S.~D. Katz, and C.~Schmidt, {\it {The density of states method at
  non-zero chemical potential}},  {\em JHEP} {\bf 03} (2007) 121,
  [\href{http://arxiv.org/abs/hep-lat/0701022}{{\tt hep-lat/0701022}}].

\bibitem{DElia:2007bkz}
M.~D'Elia, F.~Di~Renzo, and M.~P. Lombardo, {\it {The strongly interacting
  quark gluon plasma, and the critical behaviour of QCD at imaginary $\mu$}},
  {\em Phys. Rev.} {\bf D76} (2007) 114509,
  [\href{http://arxiv.org/abs/0705.3814}{{\tt arXiv:0705.3814}}].

\bibitem{Li:2010qf}
A.~Li, A.~Alexandru, K.-F. Liu, and X.~Meng, {\it {Finite density phase
  transition of QCD with $N_f=4$ and $N_f=2$ using canonical ensemble method}},
   {\em Phys. Rev.} {\bf D82} (2010) 054502,
  [\href{http://arxiv.org/abs/1005.4158}{{\tt arXiv:1005.4158}}].

\bibitem{Endrodi:2018zda}
G.~Endrodi, Z.~Fodor, S.~D. Katz, D.~Sexty, K.~K. Szabo, and C.~Torok, {\it
  {Applying constrained simulations for low temperature lattice QCD at finite
  baryon chemical potential}},  {\em Phys. Rev.} {\bf D98} (2018), no.~7
  074508, [\href{http://arxiv.org/abs/1807.08326}{{\tt arXiv:1807.08326}}].

\bibitem{Ohnishi:2015fhj}
A.~Ohnishi, {\it {Approaches to QCD phase diagram; effective models,
  strong-coupling lattice QCD, and compact stars}},  {\em J. Phys. Conf. Ser.}
  {\bf 668} (2016), no.~1 012004, [\href{http://arxiv.org/abs/1512.08468}{{\tt
  arXiv:1512.08468}}].

\bibitem{Splittorff:2014zca}
K.~Splittorff, {\it {Dirac spectrum in complex Langevin simulations of QCD}},
  {\em Phys. Rev.} {\bf D91} (2015), no.~3 034507,
  [\href{http://arxiv.org/abs/1412.0502}{{\tt arXiv:1412.0502}}].

\bibitem{Nagata:2016alq}
K.~Nagata, J.~Nishimura, and S.~Shimasaki, {\it {Gauge cooling for the
  singular-drift problem in the complex Langevin method - a test in Random
  Matrix Theory for finite density QCD}},  {\em JHEP} {\bf 07} (2016) 073,
  [\href{http://arxiv.org/abs/1604.07717}{{\tt arXiv:1604.07717}}].

\bibitem{PhysRevD.109.014509}
E.~Seiler, D.~Sexty, and I.-O. Stamatescu, {\it Complex langevin: Correctness
  criteria, boundary terms, and spectrum},  {\em Phys. Rev. D} {\bf 109} (Jan,
  2024) 014509.

\bibitem{Seiler:2012wz}
E.~Seiler, D.~Sexty, and I.-O. Stamatescu, {\it {Gauge cooling in complex
  Langevin for QCD with heavy quarks}},  {\em Phys. Lett.} {\bf B723} (2013)
  213--216, [\href{http://arxiv.org/abs/1211.3709}{{\tt arXiv:1211.3709}}].

\bibitem{Aarts:2009dg}
G.~Aarts, F.~A. James, E.~Seiler, and I.-O. Stamatescu, {\it {Adaptive stepsize
  and instabilities in complex Langevin dynamics}},  {\em Phys. Lett.} {\bf
  B687} (2010) 154--159, [\href{http://arxiv.org/abs/0912.0617}{{\tt
  arXiv:0912.0617}}].

\bibitem{Banks:1979yr}
T.~Banks and A.~Casher, {\it {Chiral symmetry breaking in confining theories}},
   {\em Nucl. Phys.} {\bf B169} (1980) 103--125.

\bibitem{Ueda:2014rya}
S.~Ueda, S.~Aoki, T.~Aoyama, K.~Kanaya, H.~Matsufuru, S.~Motoki, Y.~Namekawa,
  H.~Nemura, Y.~Taniguchi, and N.~Ukita, {\it {Development of an object
  oriented lattice QCD code 'Bridge++'}},  {\em J. Phys. Conf. Ser.} {\bf 523}
  (2014) 012046.

\bibitem{Hirasawa:2020bnl}
M.~Hirasawa, A.~Matsumoto, J.~Nishimura, and A.~Yosprakob, {\it {Complex
  Langevin analysis of 2D U(1) gauge theory on a torus with a $\theta$ term}},
  \href{http://arxiv.org/abs/2004.13982}{{\tt arXiv:2004.13982}}.

\bibitem{Yokota:2023osv}
T.~Yokota, Y.~Ito, H.~Matsufuru, Y.~Namekawa, J.~Nishimura, A.~Tsuchiya, and
  S.~Tsutsui, {\it {Color superconductivity on the lattice \textemdash{}
  analytic predictions from QCD in a small box}},  {\em JHEP} {\bf 06} (2023)
  061, [\href{http://arxiv.org/abs/2302.11273}{{\tt arXiv:2302.11273}}].

\bibitem{Miura:2025aaa}
K.~Miura, Y.~Asano, Y.~Ito, H.~Matsufuru, Y.~Namekawa, J.~Nishimura,
  A.~Tsuchiya, S.~Tsutsui, and T.~Yokota {\!\!, work in progress}.

\bibitem{Osborn:2005ss}
J.~C. Osborn, K.~Splittorff, and J.~J.~M. Verbaarschot, {\it {Chiral symmetry
  breaking and the Dirac spectrum at nonzero chemical potential}},  {\em Phys.
  Rev. Lett.} {\bf 94} (2005) 202001,
  [\href{http://arxiv.org/abs/hep-th/0501210}{{\tt hep-th/0501210}}].

\end{thebibliography}\endgroup

\end{document}